\documentclass[twocolumn]{IEEEtran}
\usepackage{cite}
\usepackage{amsmath,amssymb,amsfonts}
\usepackage{algorithmic}
\usepackage{graphicx}
\usepackage{epstopdf}
\usepackage{textcomp}
\usepackage{multirow}
\usepackage{booktabs}
\usepackage{subfigure}
\usepackage{stfloats}
\usepackage{color}
\usepackage{bm}
\usepackage{array}
\usepackage{float}
\usepackage{color}
\usepackage{threeparttable}
\usepackage{booktabs}
\usepackage{tablefootnote}
\def\BibTeX{{\rm B\kern-.05em{\sc i\kern-.025em b}\kern-.08em
		T\kern-.1667em\lower.7ex\hbox{E}\kern-.125emX}}

\begin{document}
	
	\title{Nonconformal Domain Decomposition Method Based on the Hybrid SIE-PDE Formulation for Flexible Transverse Magnetic Analysis}
	\author{\IEEEauthorblockN{Aipeng Sun, \IEEEmembership{Graduate Student Member, IEEE}, Zekun Zhu, \IEEEmembership{Graduate Student Member, IEEE}, \\ Shunchuan Yang, \IEEEmembership{Member, IEEE}, and Zhizhang Chen, \IEEEmembership{Fellow, IEEE} \\}
		
		\thanks{Manuscript received xxx; revised xxx.}
		\thanks{This work was supported in part by the National Natural Science Foundation of China under Grant 62141405, 62101020, 62071125, in part by Defense Industrial Technology Development Program under Grant JCKY2019601C005, in part by Pre-Research Project under Grant J2019-VIII-0009-0170 and Fundamental Research Funds for the Central Universities. {\it {(Corresponding author: Shunchuan Yang.)}}
			
			A. Sun is with the School of Electronic and Information Engineering, Beihang University, Beijing, 100083, China (e-mail: sunaipeng1997@buaa.edu.cn).
			
			Z. Zhu and S. Yang are with the Research Institute for Frontier Science and the School of Electronic and Information Engineering, Beihang University, Beijing, 100083, China (e-mail: zekunzhu@buaa.edu.cn, scyang@buaa.edu.cn).
			
			Z. Chen is currently with the College of Physics and Information Engineering, Fuzhou University, Fuzhou, Fujian. P. R. China, on leave from the Department of Electrical and Computer Engineering, Dalhousie University, Halifax, Nova Scotia, Canada B3H 4R2 (email: zz.chen@ieee.org).
		}
	}
	
	\maketitle
	
	\begin{abstract}	
		A nonconformal domain decomposition method based on the hybrid surface integral equation partial differential equation (SIE-PDE) formulation is proposed to solve the transverse magnetic electromagnetic problems. In the hybrid SIE-PDE formulation, an equivalent model with only the electric current density is first constructed, and then is embedded into the inhomogeneous Helmholtz equation as an excitation. A connection matrix, which couples the interfaces of the SIE and PDE domains, is carefully designed to support nonconformal meshes. Since meshes in each domain are independently generated, it is much more efficient and flexible to model multiscale and complex structures compared with the original hybrid SIE-PDE formulation with conformal meshes. The proposed formulation is efficient, flexible and easy to implement. Its accuracy, efficiency and flexibility are validated by three numerical examples. 
	\end{abstract}
	

	\section{Introduction}
	\IEEEPARstart{H}{ybrid} methods gained much attention in the last few decades, since the merits of each method can possibly be used in solving challenging electromagnetic problems. One main group of hybrid methods in the frequency domain is to combine various formulations, and numerous efforts have been made, such as the method of moment/geometrical optics (MoM/PO) method [\citen{DjordMoM-PO2005}-\citen{LiuMoM-PO2012}], the method of moment uniform geometrical theory of diffraction (MoM/UTD) method [\citen{ThieleMOM-UTD1975}-\citen{TapMoM-UTD2005}], the finite-element
	method/method of moment (FEM/MoM) [\citen{AliFEM/MoM1997}-\citen{RenFEM/MoM2016}], the finite element boundary integral (FE-BI) formulation [\citen{ShengFE/BI1998}-\citen{YangFE-BI-MLFMA2013}], the hybrid surface-integral-equation/finite-element
	method (SIE/FEM) formulation [\citen{Guanmultisolver2017}][\citen{Guanmultisolver2016}], and so on. In those formulations, an artificial boundary condition, such as the third kind of boundary condition [\citen{Jin2015FEM}, Ch, 1, pp. 42-46], or the transmission condition [\citen{Jin2015FEM}, Ch. 14, pp. 1066-1083], is used to couple different formulations. Those boundary conditions are under certain assumption, which may lead to accuracy degeneration in near field calculations, like skin effect in high frequency regions. 
	
	To mitigate this issue, a hybrid surface integral equation partial differential equation (SIE-PDE) formulation is proposed in [\citen{SunSIE-PDE2022}]. In this formulation, the single-source integral equation (SS-SIE) formulation is used to construct an equivalent model for piecewise homogeneous media, and only the surface equivalent electric current density is derived. Then, it is introduced into the inhomogeneous Helmholtz equation as an additional excitation. Therefore, no additional boundary conditions, such as the Robin transmission boundary condition, are required. It can significantly improve the efficiency over the traditional FEM. However, conformal meshes should be used between the SIE and PDE domains, which inevitably limits its flexibility to handle complex structures. 
	
	In essence, the aforementioned hybrid SIE-PDE formulation can be cast into conformal domain decomposition methods (DDMs) [\citen{StupfelDDM1996}-\citen{StupfelDDM2000}], which are constrained by the mesh-conformal requirement on the interfaces of different domains. As a result, each domain cannot be meshed independently, which limits the flexibility and degenerates the efficiency. To overcome this issue, many nonconformal DDMs have been developed over the past years, such as nonconformal FEM-DDMs [\citen{LueDDM2019}][\citen{XuehybridDDM2014}], nonconformal SIE-DDMs [\citen{PengNCDDM2013}][\citen{ZhuSS-SIE2022}], nonconformal FE-BI DDMs [\citen{GaoFETI-DPDDM2016}], and nonconformal finite element method boundary element method (FEM-BEM) DDMs [\citen{RenSIE-FEM-SEM2017}][\citen{YangFEM-BEM-DDM2021}]. In general, nonconformal meshes on the interfaces are mostly handled based on the transmission conditions, such as the Dirichlet transmission condition [\citen{GaoFETI-DPDDM2016}], the second-order transmission condition [\citen{XuehybridDDM2014}][\citen{YangFEM-BEM-DDM2021}], and the Robin-type transmission condition [\citen{LueDDM2019}][\citen{PengNCDDM2013}]. However, due to these boundary conditions, these decomposed subproblems need to be solved both locally within each domain, and iteratively between domains, which may lead to efficiency issues if interfaces are not properly handled.
	
	This paper aims to develop an efficient approach to alleviate requirements upon conformal meshes in the hybrid SIE-PDE formulation without boundary conditions [\citen{SunSIE-PDE2022}], which can overcome issues in existing hybrid methods and DDMs. Based on our previous SIE-PDE formulation with conformal meshes [\citen{SunSIE-PDE2022}], the computational domain is divided into two domains: the SIE domain and the PDE domain. In the SIE domain, an equivalent model with only the electric current density is constructed for piecewise homogeneous media, where multiscale structures, and computationally challenging media, such as highly conductive media in high frequency region, are included. Then, the surface equivalent current density is enforced into the Helmholtz equation to couple the SIE and PDE formulations. Unlike the original SIE-PDE formulation, in which conformal meshes are used, a connection matrix is carefully constructed to couple nonconformal meshes between the SIE and PDE domains. Therefore, the proposed hybrid formulation is much more flexible than our previous work in [\citen{SunSIE-PDE2022}].
	
	Compared with other existing and our previous work in [\citen{SunSIE-PDE2022}], the contributions of this paper are mainly three aspects. 
	\begin{enumerate}
		\item A hybrid SIE-PDE formulation with nonconformal meshes is proposed to accurately and efficiently solve the electromagnetic problems. Since nonconformal meshes are supported, it significantly extends the flexibility of our previous work to model multiscale and complex structures. 
		
		\item Two special scenarios are considered in the proposed formulations. The original hybrid SIE-PDE formulation can be obtained from the proposed SIE-PDE formulation with nonconformal meshes. 
		
		\item Several numerical examples including coated dielectric objects, and skin effect in cables, are carried out to validate the accuracy, efficiency and flexibility of the proposed hybrid formulation with nonconformal meshes. 
	\end{enumerate}	

	This paper is organized as follows. In Section II, a brief review of the hybrid SIE-PDE formulation with conformal meshes is presented. In Section III, the proposed hybrid SIE-PDE formulation with nonconformal meshes are demonstrated in detail. In Section IV, two special scenarios of the proposed SIE-PDE formulation are considered. Then, three numerical examples are carried out to validate its accuracy, efficiency, and flexibility. Finally, we draw some conclusions in Section V.

	\section{Review of the Hybrid SIE-PDE Formulation}
	Before the proposed formulation is presented, a brief review of the original hybrid SIE-PDE formulation in [\citen{SunSIE-PDE2022}] is first presented. 
	
\begin{figure}
		\begin{minipage}[t]{0.25\textwidth}
			\centering
			\centerline{\includegraphics[scale=0.24]{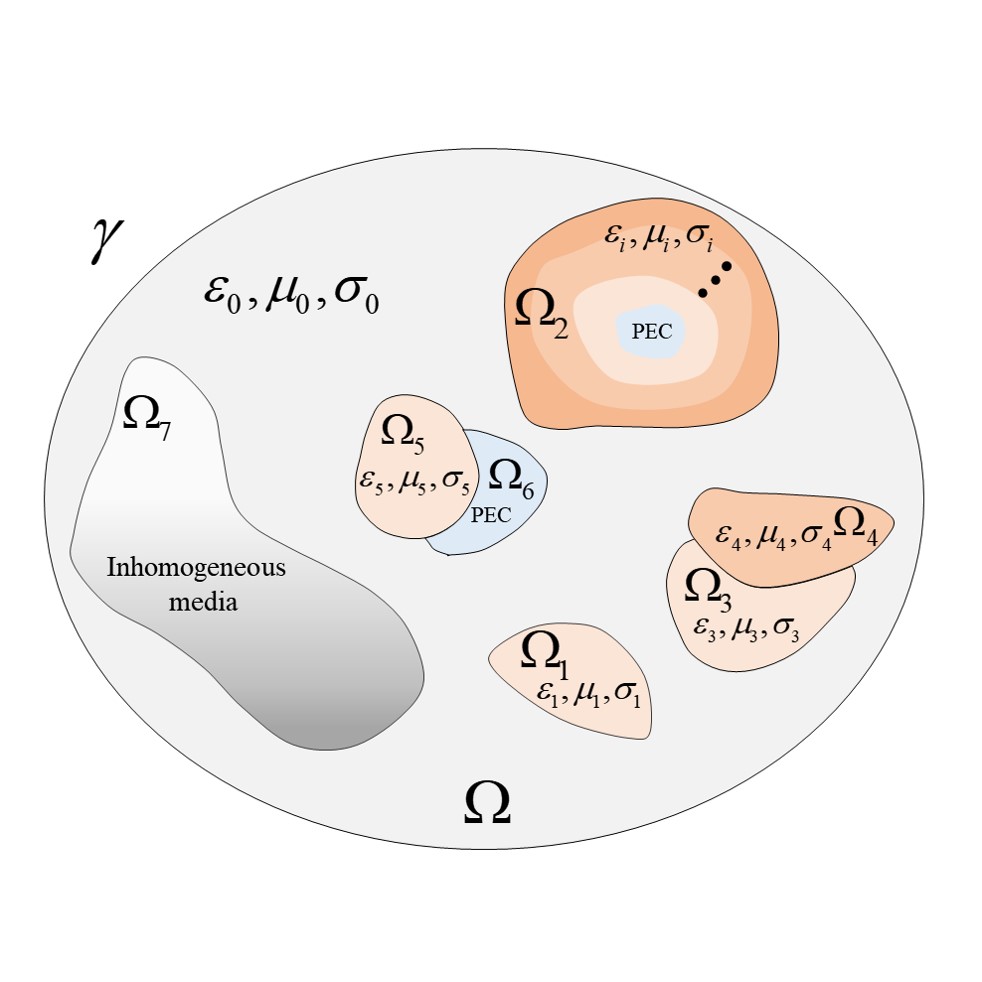}}
			\centering
			\centerline{(a)}
		\end{minipage}
		\begin{minipage}[t]{0.23\textwidth}
			\centering
			\centerline{\includegraphics[scale=0.24]{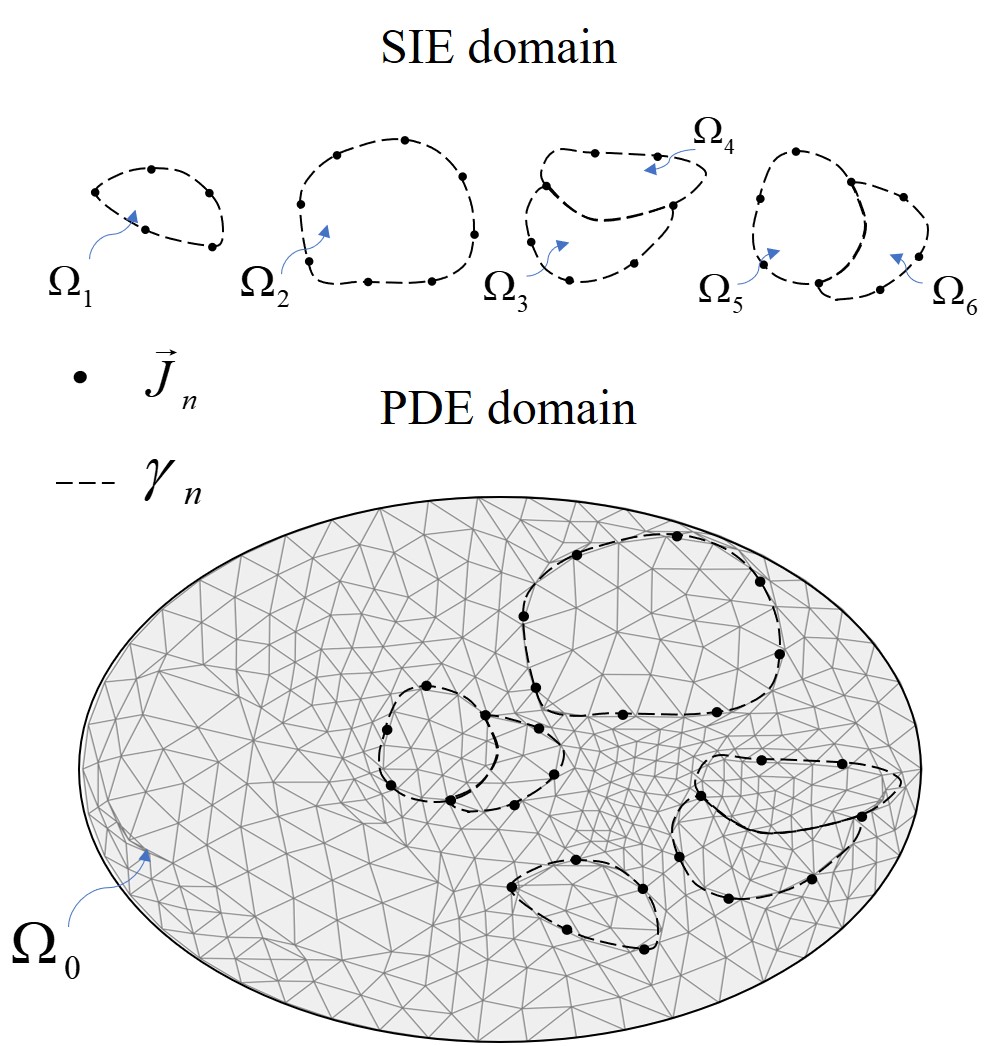}}
			\centering
			\centerline{(b)}
		\end{minipage}
		\caption{(a) The general scenario including complex structures with homogeneous and inhomogeneous media, and (b) the overlapping SIE and PDE domains for the proposed hybrid SIE-PDE formulation. (from [\citen{SunSIE-PDE2022}]).}
		\label{general_model}
	\end{figure}

	As shown in Fig. \ref{general_model}, the whole computational domain is first decomposed into the overlapping SIE and PDE domains. In the SIE domain, complex structures with piecewise homogeneous media are included. According to the surface equivalence theorem (SET) [\citen{LOVE}, Ch.12, pp. 653-658], those structures are replaced by the homogeneous background medium, and an equivalent model with only the surface equivalent electric current density on an enclosed contour is derived to represent their electromagnetic effects. The PDE domain contains the remaining computational region and the region, where structures are replaced by the background medium. Through substituting the equivalent current density into the inhomogeneous Helmholtz equation as an excitation, the hybrid SIE-PDE formulation is derived.
	  
	\subsection{The Derivation of the Hybrid SIE-PDE Formulation}  
	The electric fields $\mathbf{E}_0$ in the PDE domain satisfy the inhomogeneous Helmholtz equation. Under the two-dimensional transverse magnetic (TM) assumption, it can be expressed as
	\begin{equation}\label{Inhomo-PDE}
		\frac{\partial }{{\partial x}}\left( {\frac{1}{{{\mu _r}}}\frac{{\partial {{\bf{E}}_0}}}{{\partial x}}} \right){\rm{ + }}\frac{\partial }{{\partial y}}\left( {\frac{1}{{{\mu _r}}}\frac{{\partial {{\bf{E}}_0}}}{{\partial y}}} \right){\rm{ + }}k_0^2{\varepsilon _r}{{\bf{E}}_0} = j\omega {\mu _0}{\bf{J}},
	\end{equation}
	where $\mathbf{J}$ is the electric current density inside the computational domain, $k_0$ is the wave number in the free space, $\epsilon_r$, $\mu_r$ are the relative permittivity and permeability of the medium in the PDE domain, respectively, $\mu_0$ is the permeability of free space.
	
	To clearly demonstrate the procedure to construct the equivalent model, we take $\Omega_1$ as an example. Based on the SET [\citen{LOVE}, Ch.12, pp. 653-658], the electric current density in the equivalent model can be expressed as
	\begin{equation} \label{DSAO}
		{{\mathbf{J}}_1} = {\left. {{{\mathcal Y}_{{s_1}}}{{\mathbf{E}}_1}} \right|_{{\mathbf{r}} \in {\gamma _1}}},
	\end{equation}
	where ${\left. {{{\mathbf{E}}_1}} \right|_{{\mathbf{r}} \in {\gamma _1}}}$ denotes electric fields ${{\mathbf{E}}_1}$ on $\gamma_1$, and ${\mathcal Y}_{{s_1}}$ is the differential surface admittance operator (DSAO) [\citen{Zutter2005DSAO}] defined on $\gamma_1$ in the continuous physical domain, which relates the surface equivalent electric current density ${\mathbf{J}}_1$ to electric fields ${{\mathbf{E}}_1}$ on $\gamma_1$.
	
	Once ${\mathbf{J}}_1$ is obtained, it can be treated as an excitation in the inhomogeneous Helmholtz equation to couple the SIE and PDE formulations. Then, by further substituting ${\mathbf{J}}_1$ with ${\left. {{{\cal Y}_{{s_1}}}{{\mathbf{E}}_1}} \right|_{{\mathbf{r}} \in {\gamma _1}}}$ into (\ref{Inhomo-PDE}), the hybrid SIE-PDE formulation can be derived as
	\begin{align} 
		\frac{\partial }{{\partial x}}\left( {\frac{1}{{{\mu _r}}}\frac{{\partial {\mathbf{E}_0}}}{{\partial x}}} \right){\rm{ + }}\frac{\partial }{{\partial y}}&\left( {\frac{1}{{{\mu _r}}}\frac{{\partial {\mathbf{E}_0}}}{{\partial y}}} \right)  
		\label{Jn+ysn_modified}{\rm{ + }}k_0^2{\varepsilon _r}{\mathbf{E}_0}\\ 
		& \qquad- j\omega {\mathcal{Y}_{s_1}}{\left.{{\mathbf{E}_1}} \right|_{{\mathbf{r}} \in {\gamma _1}}} = j\omega \mu \mathbf{J}.\notag
	\end{align} 
	
	It can be found that the SIE and PDE formulations are coupled through the surface equivalent electric current density, which does not require additional boundary conditions, and is mathematically equivalent to the original physical model. 
	
		\begin{figure}
		\begin{minipage}[t]{0.25\textwidth}
			\centering
			\centerline{\includegraphics[scale=0.25]{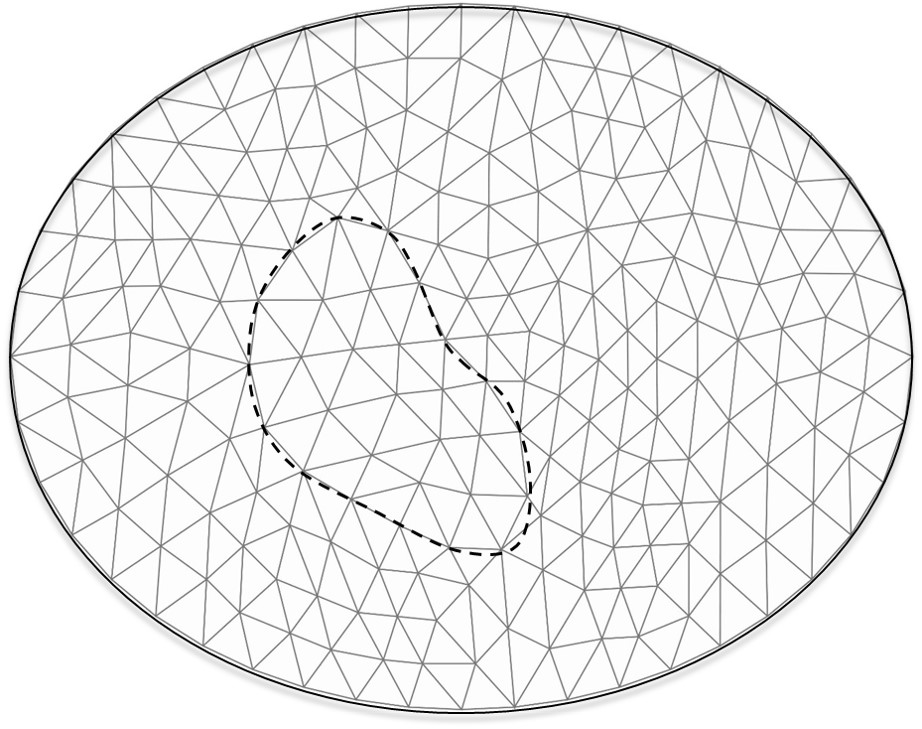}}
			\centering
			\centerline{(a)}
		\end{minipage}
		\begin{minipage}[t]{0.23\textwidth}
			\centering
			\centerline{\includegraphics[scale=0.25]{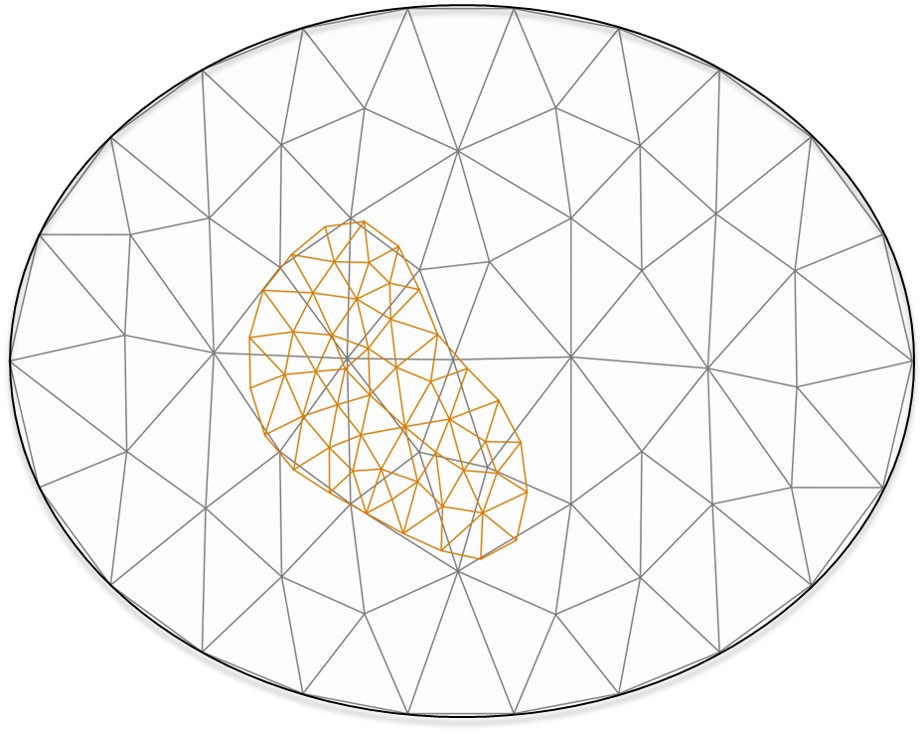}}
			\centering
			\centerline{(b)}
		\end{minipage}
		\caption{(a) The original conformal meshes, and (b) the nonconformal meshes in the whole computational domain.}
		\label{mesh_comparison}
	\end{figure}
	
	\subsection{Numerical Implementation of the Hybrid SIE-PDE Formulation}
	In the practical implementation, the SIE and PDE formulations are accurately coupled at the interface through appropriate basis functions. The rooftop basis function is used to expand electric fields in the SIE domain, and the Galerkin scheme is used for testing. Then, $\mathbf{J}_1$ can be expressed as
	\begin{equation} \label{Discrete-DSAO}
		{\mathbf{J}_1} = {{\mathbb{Y}}_{s_1}}{\mathbf{E}_1},
	\end{equation}
	where $\mathbf{J}_1$, $\mathbf{E}_1$ are two column vectors composed of their expansion coefficients, and ${\mathbb{Y}}_{s_1}$ is the discrete DSAO, which corresponds to ${\mathcal Y}_{{s_1}}$ in (\ref{DSAO}).
	
	In the PDE domain, the linear basis functions are selected to discretize electric fields, and the Galerkin scheme is used to test the hybrid SIE-PDE formulation. By assembling all the elemental equations, and enforcing that the residual error vanishes, the final matrix equation can be obtained as
	\begin{equation} \label{pre-SIE-PDE}
		{\mathbb{K}}{\mathbf{E}} + {\mathbb{B}}{\mathbf{J}_1} = {\mathbf{g}} - {\mathbf{b}}.
	\end{equation}
	By substituting (\ref{Discrete-DSAO}) into (\ref{pre-SIE-PDE}), and replacing $\mathbf{E}_1$ with $\mathbf{E}$, it can be further rewritten as
	\begin{equation} \label{SIE-PDE}
		\left( {\mathbb{K}} + {\mathbb{A}} \right) {\mathbf{E}}= {\mathbf{g}} - {\mathbf{b}},
	\end{equation}
	where $\mathbb{A}$ denotes the expanded matrix for ${\mathbb{B}}{\mathbb{Y}}_{s_1}$ . Through solving (\ref{SIE-PDE}), we can obtain electric fields in the PDE domain. Detailed expressions of matrices are given in [\citen{SunSIE-PDE2022}]. 
	
	It can be found that the SIE-PDE formulation can accurately couple the SIE and PDE formulations through introducing the surface equivalent current densities. When the computationally challenging media including highly conductive interconnects and complex structures are considered, the original SIE-PDE formulation can be computationally efficient. However, one implication in the SIE-PDE formulation is that meshes on the interfaces between the SIE and PDE formulations require to be conformal, which inevitably reduces its flexibility to model complex structures. In this article, such constraint is removed through introducing nonconformal meshes and further enhances the flexibility of the SIE-PDE hybrid formulation. We reported the preliminary idea in [\citen{SunNCSIE-PDE2022}]. Detailed formulations and discussion will be presented in the following sections.
	
	\section{The Proposed Hybrid SIE-PDE Formulation with Nonconformal Meshes}
	\subsection{Node Transformation from the SIE Domain to the PDE Domain}
	 As shown in Fig. \ref{mesh_comparison} (b), compared with Fig. \ref{mesh_comparison} (a), nonconformal meshes are used to discretize the SIE and PDE formulations separately. By considering different material constants in each domain, the corresponding mesh densities can be used. In Fig. \ref{mesh_comparison} (b), fine meshes are used in the SIE domain, and coarse meshes are applied in the PDE domain for illustration.
	 
	 As discussed in Section II-D, the hybrid SIE-PDE formulation is constructed based on the PDE formulation. We define a connection matrix $\mathbb{T}$, which interpolates field values from the triangular element in the PDE domain to interface nodes on the boundary of the SIE domain.
	 
	 Assuming that there are $m$ nodes on the boundary of the SIE domain, which can be transformed into $m_s$ nodes in the PDE domain, the relationship among the expansion coefficients defined on these nodes can be expressed via $\mathbb{T}$ as
	 \begin{equation} \label{Expand_via_T}
	 	\mathbf{E_1} = \mathbb{T}{\widetilde{\mathbf{E}}_1}
	 \end{equation}
 	where
 	\begin{align}
 		&\mathbf{E_1} = {\left[ {\begin{array}{*{20}{c}}
 					{{e_1}}&{{e_2}}& \ldots &{{e_m}}
 			\end{array}} \right]^T}, \notag \\
 		&{\widetilde{\mathbf{E}}_1} = {\left[ {\begin{array}{*{20}{c}}
 					{{{\tilde e}_1}}&{{{\tilde e}_2}}& \ldots &{{{\tilde e}_{{m_s}}}}
 			\end{array}} \right]^T}.\notag
 	\end{align}
 	${e_i}\left( {i = 1,2,...,m} \right)$ and ${\tilde e_j}\left( {j = 1,2,...,{m_s}} \right)$ are expansion coefficients of $\mathbf{E_1}$ before and after the interpolation, respectively. $\mathbb{T}$ is a connection matrix with dimension of $m \times {m_s}$. In the next subsection, details of its construction are presented.
 	
 	\subsection{Construction of the Connection Matrix $\mathbb{T}$}
\begin{figure}[t]
 		\centering
 		\includegraphics[width=0.31\textwidth]{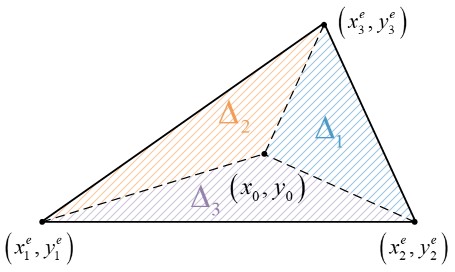}
 		\caption{ ${\Delta _j}\left( {j = 1,2,3} \right)$ within a triangular element.}
 		\label{triangle} 
 	\end{figure}

 	As the FEM analysis, the linear interpolation functions can be used to construct the connection matrix $\mathbb{T}$. The interpolation functions on the $e$th triangular element [\citen{Jin2015FEM}] are defined as
 	\begin{equation} \label{Interpolation_function}
 		\begin{array}{*{20}{c}}
 			{L_j^e\left( {x,y} \right) = \displaystyle \frac{{{\Delta _j}}}{{{\Delta ^e}}} = \frac{1}{{2{\Delta ^e}}}\!\left( {a_j^e + b_j^ex + c_j^ey} \right),}&{\!\!\!\!\!j = 1,2,3,}
 		\end{array}
 	\end{equation}   
 	where 
 		\[\begin{array}{*{20}{l}}
 		{a_1^e = x_2^ey_3^e - y_2^ex_3^e}&{b_1^e = y_2^e - y_3^e}&{c_1^e = x_3^e - x_2^e},\\
 		{a_2^e = x_3^ey_1^e - y_3^ex_1^e}&{b_2^e = y_3^e - y_1^e}&{c_2^e = x_1^e - x_3^e},\\
 		{a_3^e = x_1^ey_2^e - y_1^ex_2^e}&{b_3^e = y_1^e - y_2^e}&{c_3^e = x_2^e - x_1^e},
 	\end{array}\]
 	and
 	\[{\Delta ^e} = \frac{1}{2}\left| {\begin{array}{*{20}{c}}
 			1&{x_1^e}&{y_1^e}\\
 			1&{x_2^e}&{y_2^e}\\
 			1&{x_3^e}&{y_3^e}
 	\end{array}} \right| = \frac{1}{2}\left( {b_1^ec_2^e - b_2^ec_1^e} \right).\]
 	$x_i^e$ and $y_i^e$ ($i = 1,2,3$) denote the coordinates in the $x$, $y$ direction of the $i$th node of the $e$th triangle element, respectively, and ${\Delta ^e}$ denotes the area of the $e$th element. The definitions of ${\Delta _j}$ ($j = 1,2,3$) within a triangular element are shown in Fig. \ref{triangle}.   
	In order to construct $\mathbb{T}$, the following three steps are generally needed:
	\begin{enumerate}
		\item Determine in which triangular element the boundary nodes of the SIE domain are located in the PDE domain. It can be done through the relationship between the interpolation functions defined in the $e$th triangular element, given by
		\begin{equation} \label{interpolation_function_relation}
			\begin{array}{*{20}{c}}
				{\sum\limits_{j = 1}^3 {L_j^e\left( {{x_0},{y_0}} \right)}  = 1,}&{L_j^e\left( {{x_0},{y_0}} \right) \in \left[ {0,1} \right]}
			\end{array}.
		\end{equation}
		$x_0$ and $y_0$ denote the coordinates in the $x$, $y$ direction of the boundary nodes in the SIE domain, respectively. If (\ref{interpolation_function_relation}) is satisfied, the triangle element can then be determined, and the corresponding interpolation coefficients can be denoted by $l_j^e$ ($j = 1,2,3$).
		\item After all the triangular elements including the boundary nodes of the SIE domain are determined, the global numbers of their nodes in the PDE domain can be derived. Assuming $m_s$ nodes are involved, the relationship from their local indexes $k$ ($k = 1, \cdots ,{m_s}$) to the related global counterparts $M\left( k \right)$ can then be obtained, where $M\left(  \cdot  \right)$ denotes the mapping function.
		\item The connection matrix $\mathbb{T}$ can be constructed based on the mapping function. For the $i$ ($i = 1, \cdots ,m$)th nodes on the boundary of the SIE domain, it is located in the $e^i$th triangular element with nodes $M\left( {e_1^i} \right)$, $M\left( {e_2^i} \right)$, $M\left( {e_3^i} \right)$ in the PDE domain. Therefore, entities of $\mathbb{T}$ are given by
		\begin{equation}
			\begin{array}{*{20}{c}}
				{{{\left[ {\mathbb{T}} \right]}_{i,e_j^i}} = l_j^{{e^i}},}&{j = 1,2,3}
			\end{array}.
		\end{equation} 
		\end{enumerate} 
	   
		Therefore, $\mathbb{T}$ can be constructed.
		
		\subsection{The Proposed Nonconformal Hybrid SIE-PDE Formulation}
		Once $\mathbb{T}$ is constructed, it can be used to derive the proposed hybrid SIE-PDE formulation with nonconformal meshes. By substituting (\ref{Discrete-DSAO}) and (\ref{Expand_via_T}) into (\ref{pre-SIE-PDE}), the original hybrid formulation can be rewritten as
		\begin{equation} \label{pre-nonconformal-SIE-PDE}
			{\mathbb{K}}{\mathbf{E}} + {\mathbb{B}}{{\mathbb{Y}}_{s_1}}\mathbb{T}{\widetilde{\mathbf{E}}_1} = {\mathbf{g}} - {\mathbf{b}}.
		\end{equation}
		It can be found that ${\widetilde{\mathbf{E}}_1}$ is only part of ${\mathbf{E}}$. Therefore, (\ref{pre-nonconformal-SIE-PDE}) can be further rewritten as
		\begin{equation} \label{nonconformal-SIE-PDE}
			\left( {\mathbb{K}} + {\mathbb{C}} \right) {\mathbf{E}}= {\mathbf{g}} - {\mathbf{b}},
		\end{equation}
		where $\mathbb{C}$ is the expansion matrix for ${\mathbb{B}}{{\mathbb{Y}}_{s_1}}\mathbb{T}$. The proposed hybrid SIE-PDE formulation with nonconformal meshes is obtained. Through solving (\ref{nonconformal-SIE-PDE}), electric fields in the PDE domain can be obtained. However, since the SET is used in the SIE-PDE formulation, electric fields inside the SIE domain cannot be derived directly and should be carefully recovered. It will be discussed in the next subsection.
		
		\subsection{Fields Calculation in the Whole Computational Domain}
		Due to the relationship between ${\widetilde{\mathbf{E}}_1}$ and ${\mathbf{E}}$, ${\widetilde{\mathbf{E}}_1}$ can be accurately extracted from solutions of (\ref{nonconformal-SIE-PDE}), and then ${\mathbf{E}_1}$ can be derived through (\ref{Expand_via_T}). Therefore, the boundary values in the SIE domain can be obtained. After that, the electric fields inside the SIE domain can be recovered. In addition, other interested parameters, such as the surface equivalent electric current density, and the radar cross section (RCS), can also be calculated based on [\citen{ZhuSS-SIE2022}][\citen{Zhou2021embedded}][\citen{Zhou2021SS-SIE}].
		
		It should be noted that although we restrict the discussion to one PDE domain with a single SIE domain, the proposed formulation can be easily generalized for multiple SIE domains in the PDE domain as well. With the application of SET, the single-source equivalent models with only the equivalent electric current densities for $\Omega_i$ (${i = 1,2, \cdots ,5}$) in Fig. \ref{general_model} can be obtained [\citen{ZhuSS-SIE2022}][\citen{Zhou2021embedded}][\citen{Zhou2021SS-SIE}]. After that, the nonconformal hybrid SIE-PDE formulation can be derived by combining these equivalent sources.
		
		\section{Two Special Scenarios}
\begin{figure}
			\centering
			\begin{minipage}[t]{0.48\linewidth}
				\centering
				\centerline{\includegraphics[scale=0.33]{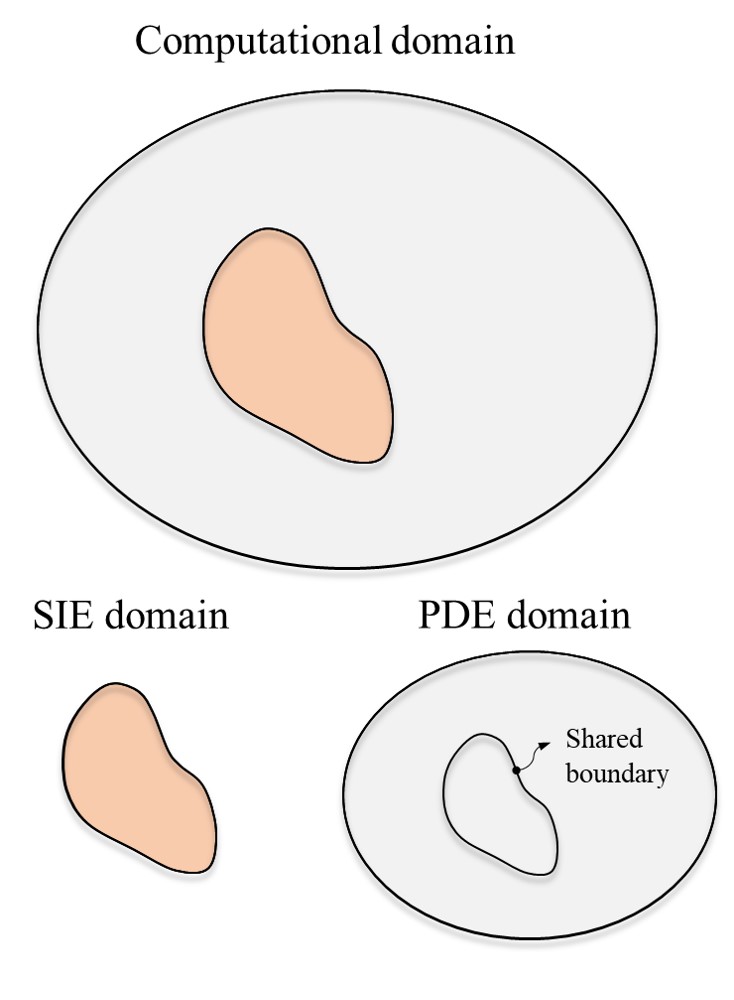}}
				\centering
				\centerline{(a)}
			\end{minipage}
			\hfill
			\begin{minipage}[t]{0.48\linewidth}
				\centering
				\centerline{\includegraphics[scale=0.33]{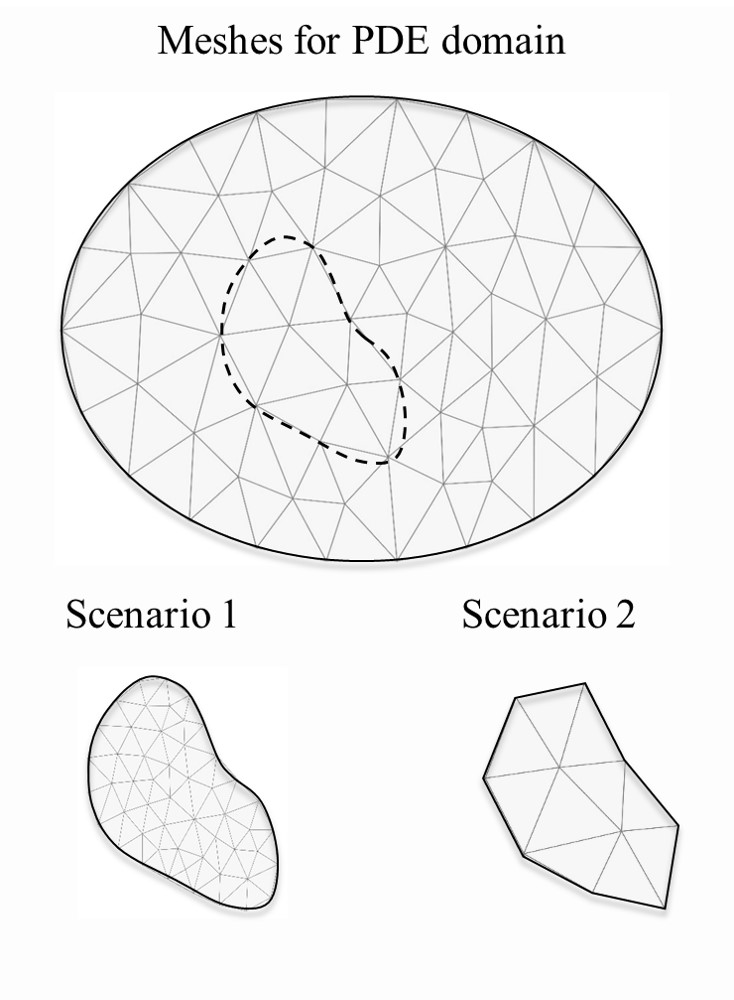}}
				\centering
				\centerline{(b)}
			\end{minipage}
			\caption{ (a) The decomposed SIE and PDE domains with a common boundary, and (b) meshes for SIE and PDE domains with two special scenarios.}
			\label{special}
		\end{figure}

		When the SIE and PDE domains share common boundaries, as shown in Fig. \ref{special} (a), the nonconformal hybrid SIE-PDE formulation would degenerate into two special scenarios. One is an incomplete nonconformal case, in which the two domains are discretized with different mesh sizes on both sides of the shared boundaries. The other is the conformal case. The two domains are the same within overlapping domains. The meshes of the two scenarios are shown in Fig. \ref{special} (b). In this section, these two special scenarios are discussed in detail.
		
		\subsection{The Incomplete Nonconformal Scenario of the Proposed Nonconformal Hybrid Formulation}
		In the first scenario, although the two domains share the same boundary, nodes on the boundary of the SIE domain would not all fall on the edges of triangular elements, since relative fine meshes are applied in the SIE domain, as shown in Fig. \ref{special} (b). In other words, the nodes may fall on the interior, edges or vertices of the triangular elements near the shared boundary. It is the same with the nonconformal hybrid SIE-PDE formulation illustrated above. Therefore, under this condition, it can also be solved following the procedures presented in Section-III.
		
		It should be noted that when nodes fall on the edges of triangular elements, they are interpolated based on line segments, and only two of the corresponding expansion coefficients, denoted as $l_j^s$ ($j = 1,2$), would be non-zero, which can be calculated by the linear interpolation functions of one-dimensional form, given by
		\begin{equation} \label{1d_interpolation_function}
			\begin{array}{*{20}{c}}
				{l_1^s} = {\displaystyle \frac{{\left| {{\mathbf{r}} - {{\mathbf{r}}_1}} \right|}}{{{l^s}}},}&{l_2^s{\rm{ = }} \displaystyle \frac{{\left| {{{\mathbf{r}}_2} - {\mathbf{r}}} \right|}}{{{l^s}}},}&{{\mathbf{r}} \in \left[ {{{\mathbf{r}}_1},{{\mathbf{r}}_2}} \right],}
			\end{array}	
		\end{equation}
		where $\mathbf{r}_1$, $\mathbf{r}_2$ are two end points of the $s$th line segment. $l^s$ denotes its length, and $\left| {{\mathbf{r}} - {{\mathbf{r}}_1}} \right|$, $\left| {{{\mathbf{r}}_2} - {\mathbf{r}}} \right|$ denote the distance between the interpolated node and end points, respectively.	 


		Since the equivalent current density always lies on the shared boundary before and after the node transformation, it is fully equivalent to the original physical model. Therefore, it is more accurate compared with the complete nonconformal formulation proposed in Section III.
		
		\subsection{Conformal Form of the Proposed Nonconformal Hybrid Formulation}
		In the second scenario, the SIE and PDE domains completely overlap, as shown in Fig. \ref{special} (b), and the proposed nonconformal hybrid formulation degenerates into the original formulation developed in [\citen{SunSIE-PDE2022}]. In this scenario, for each determined triangular element, only one of the corresponding interpolation coefficients $l_j^e$ ($j = 1,2,3$) would be non-zero with entities as 1, and the final formulation would be the same as (\ref{SIE-PDE}).  
		
		Through the connection matrix $\mathbb{T}$, these two special scenarios can be unified in the proposed formulation with nonconformal meshes. Therefore, the proposed nonconformal hybrid formulation is a more general form of hybrid SIE-PDE formulations. 
		\begin{figure}
			\centering
			\begin{minipage}[t]{0.5\linewidth}
				\centering
				\centerline{\includegraphics[scale=0.28]{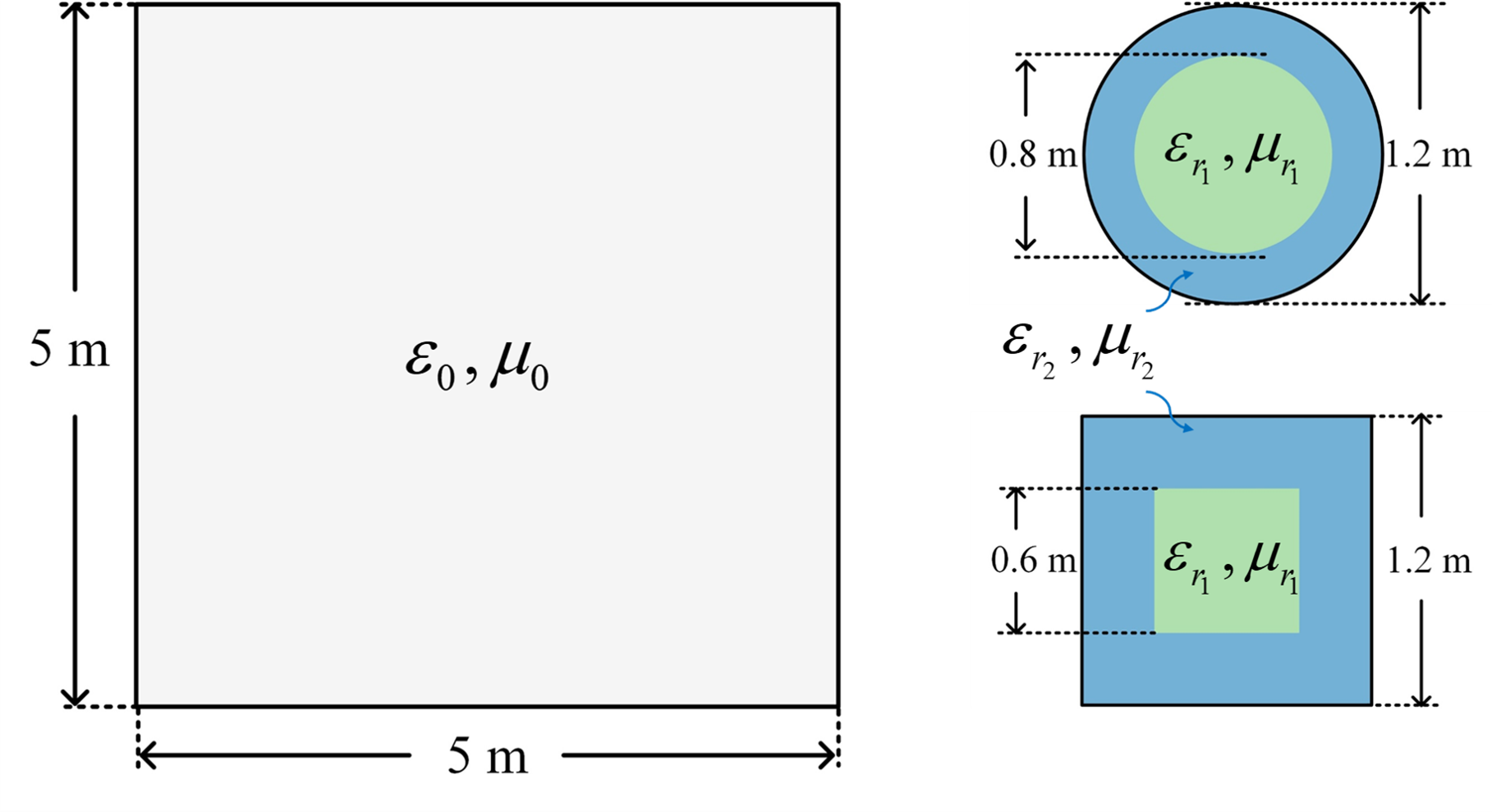}}
				\centering
				\centerline{(a)}
			\end{minipage}
			\hfill
			\begin{minipage}[t]{0.5\linewidth}
				\centering
				\centerline{\includegraphics[scale=0.26]{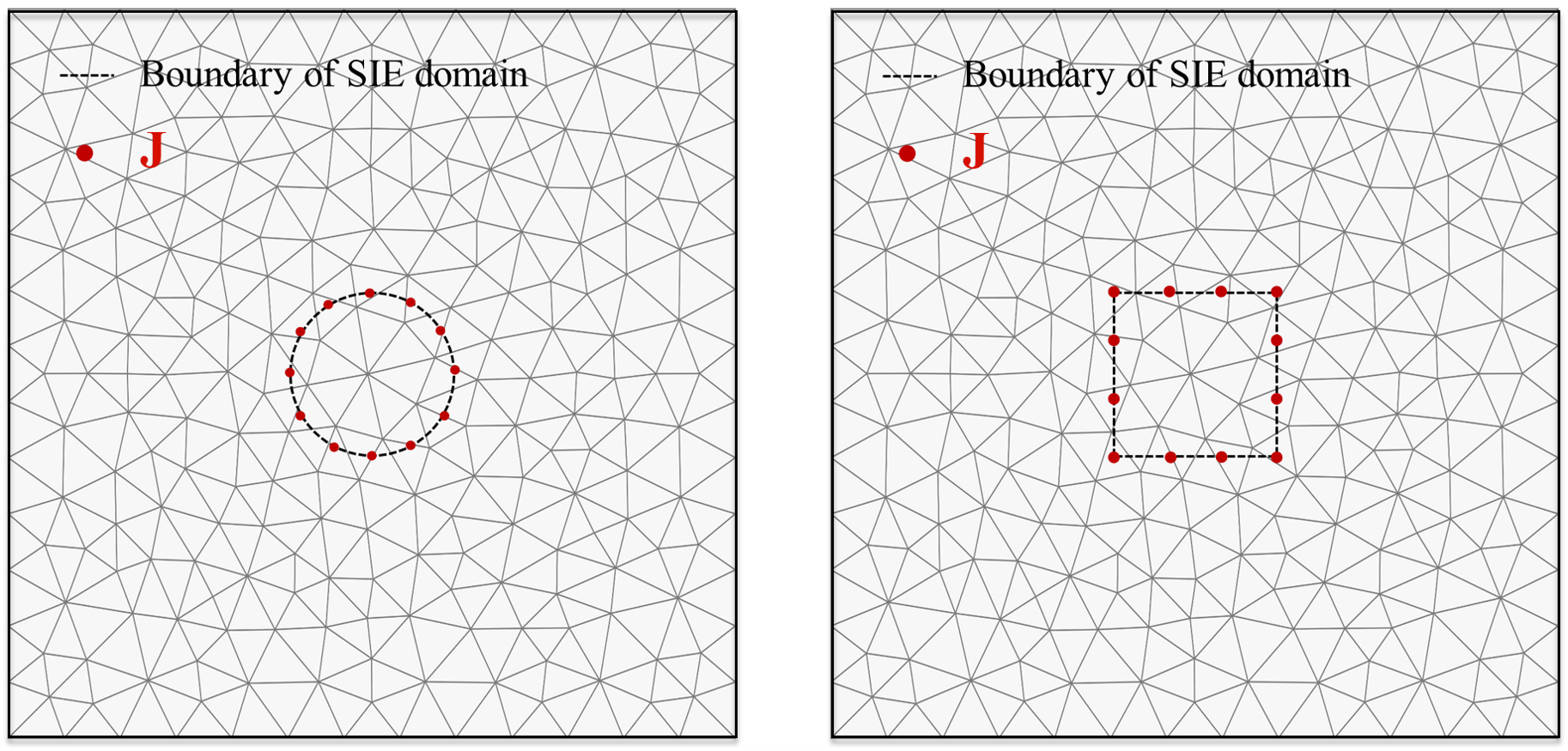}}
				\centering
				\centerline{(b)}
			\end{minipage}
			\caption{ (a) Geometrical configurations of the coated cylinder embedded in air, and (b) the nonconformal meshes of the SIE and PDE domains.}
			\label{case1_config}
		\end{figure}
		\section{Numerical Results and Discussion} 
		All numerical examples in this paper are performed on a workstation with an Intel i7-7700 3.6 GHz CPU and 64 G memory. Our in-house codes are written in Matlab. To make a fair comparison, only a single thread without any parallel computation is used to complete the simulations.

		\subsection{The Coated Dielectric Cylinders}
		In this example, an infinite long dielectric coated cylinder with circular and square cross section is considered, respectively. For the circular cross section, the radius of the inner and outer is 0.4 m and 0.6 m, and for the square cross section, the side length is 0.6 m and 1.2 m. The relative permittivity and permeability of the inner and outer dielectric object is ${\varepsilon _{{r_1}}} = 4$, ${\mu _{{r_1}}} = 1$ and ${\varepsilon _{{r_2}}} = 2.3$, ${\mu _{{r_2}}} = 1$, respectively. The whole object is embedded in air with constant parameters of $\epsilon_0$, $\mu_0$. A TM-polarized plane wave with $f=300$ MHz incidents from the $x$-axis.
		
		\begin{figure}
	\centering
	\begin{minipage}[t]{0.5\linewidth}
		\centering
		\centerline{\includegraphics[scale=0.065]{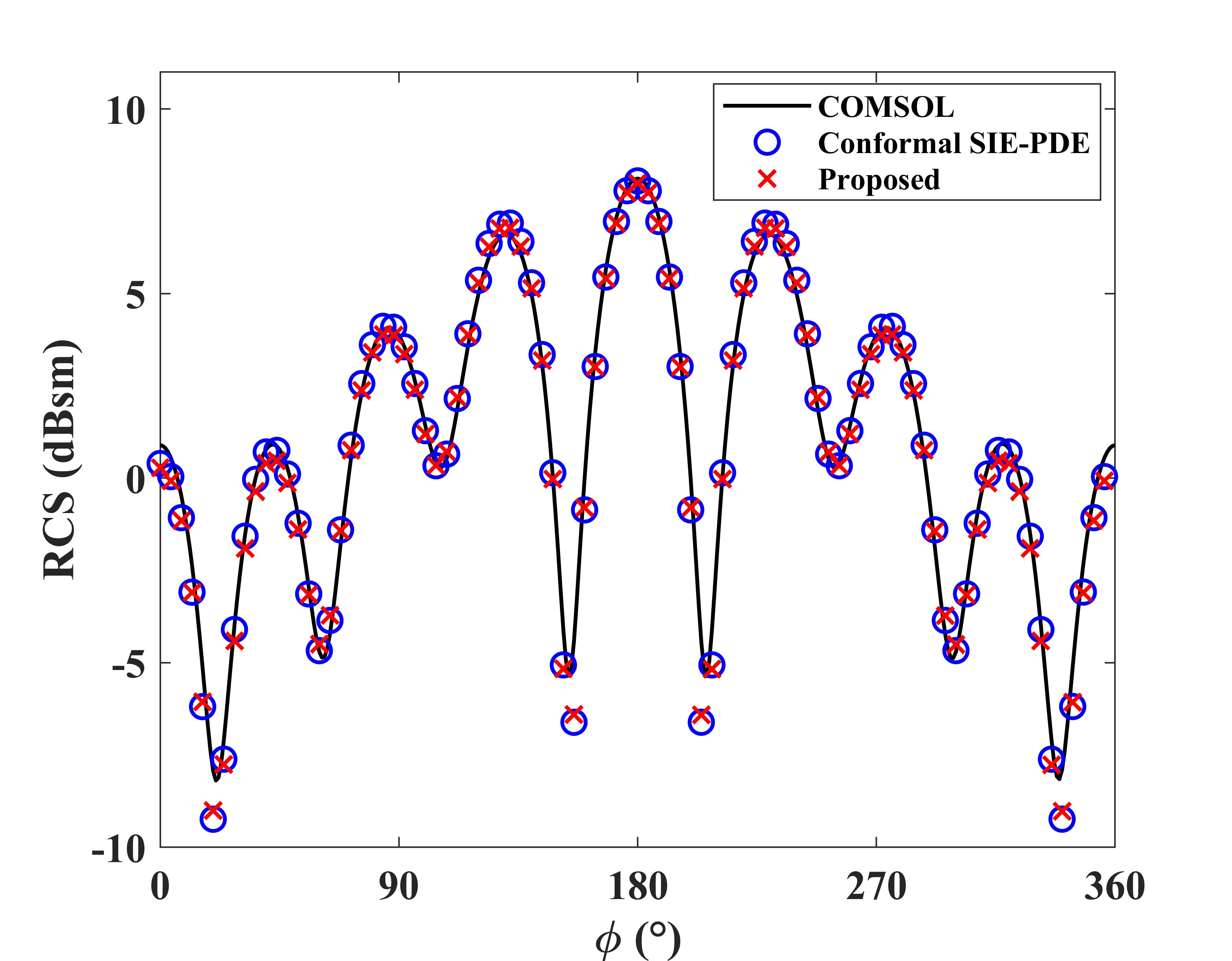}}
		\centering
		\centerline{(a)}
	\end{minipage}
	\hfill
	\begin{minipage}[t]{0.5\linewidth}
		\centering
		\centerline{\includegraphics[scale=0.065]{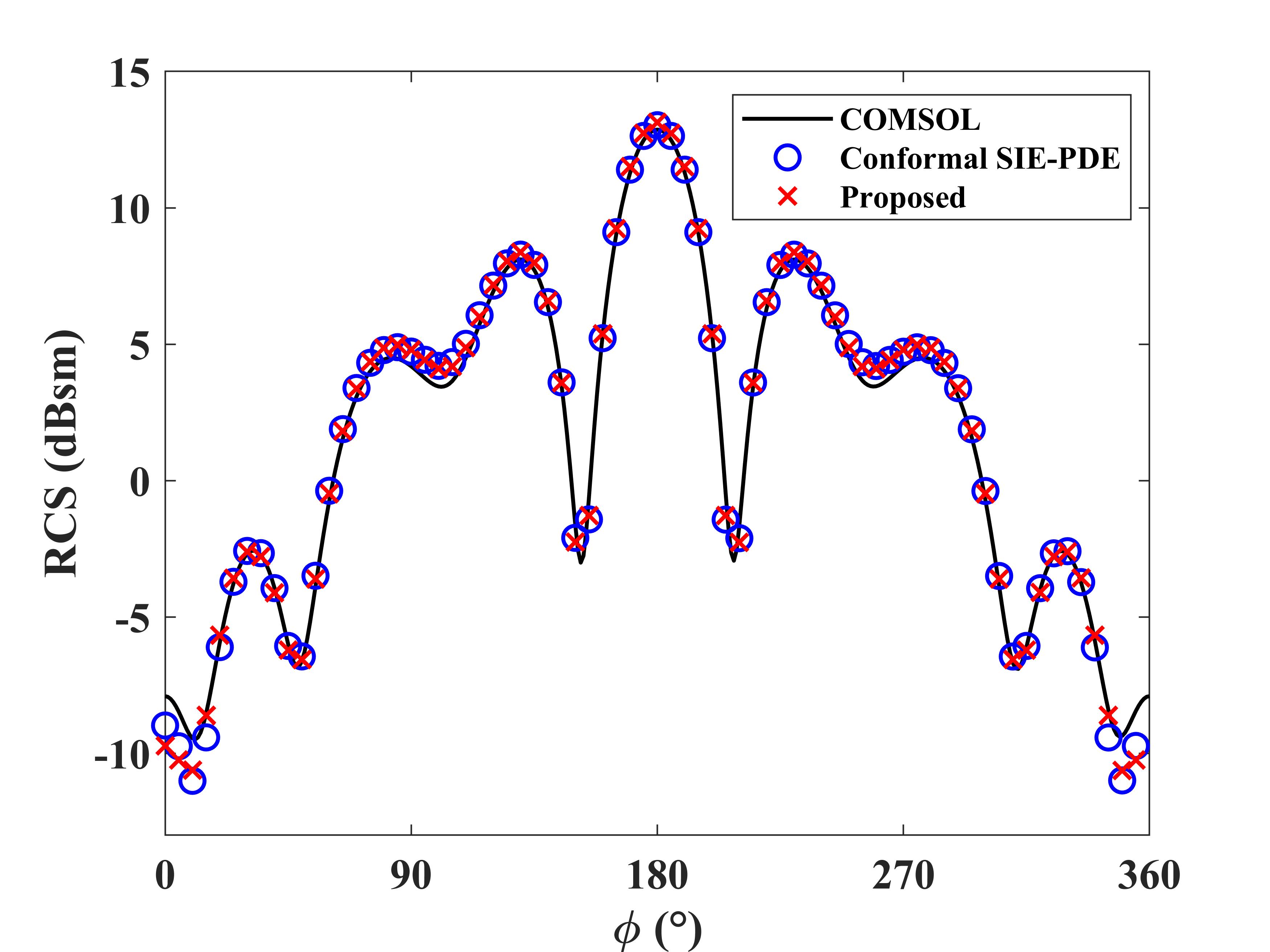}}
		\centering
		\centerline{(b)}
	\end{minipage}
	\caption{The RCS obtained from the COMSOL, the conformal hybrid SIE-PDE formulation in [\citen{SunSIE-PDE2022}], and the proposed nonconformal hybrid SIE-PDE formulation for dielectric cylinder with (a) circular cross section and (b) square cross section.}
	\label{case1_RCS}
	\end{figure}

		Through using the proposed nonconformal hybrid formulation to solve this problem, we decompose the whole computational domain into the SIE and PDE domain, as shown in Fig. \ref{case1_config} (a). The SIE domain includes the coated dielectric cylinder with an enclosed boundary, and the PDE domain is chosen as a $5m \times 5m$ airbox surrounded by 10 layers of the perfectly matched layer (PML). The nonconformal meshes for the two domains are shown in Fig. \ref{case1_config} (b). For accuracy verification, results are compared with those from the COMSOL and the conformal hybrid formulation with conformal meshes.
		
		\begin{figure}
	\centering
	\begin{minipage}[t]{0.5\linewidth}
		\centering
		\centerline{\includegraphics[scale=0.066]{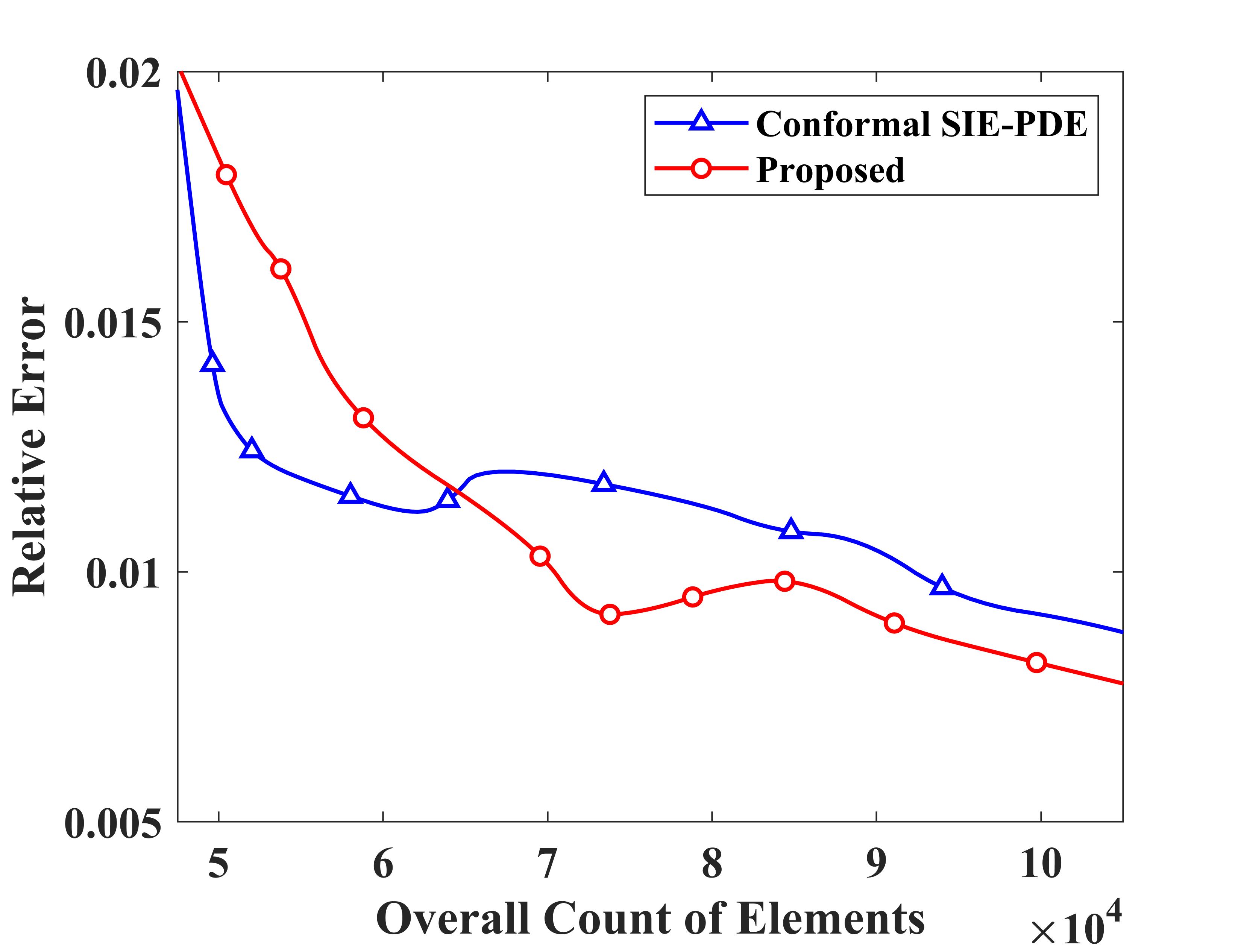}}
		\centering
		\centerline{(a)}
	\end{minipage}
	\hfill
	\begin{minipage}[t]{0.5\linewidth}
		\centering
		\centerline{\includegraphics[scale=0.061]{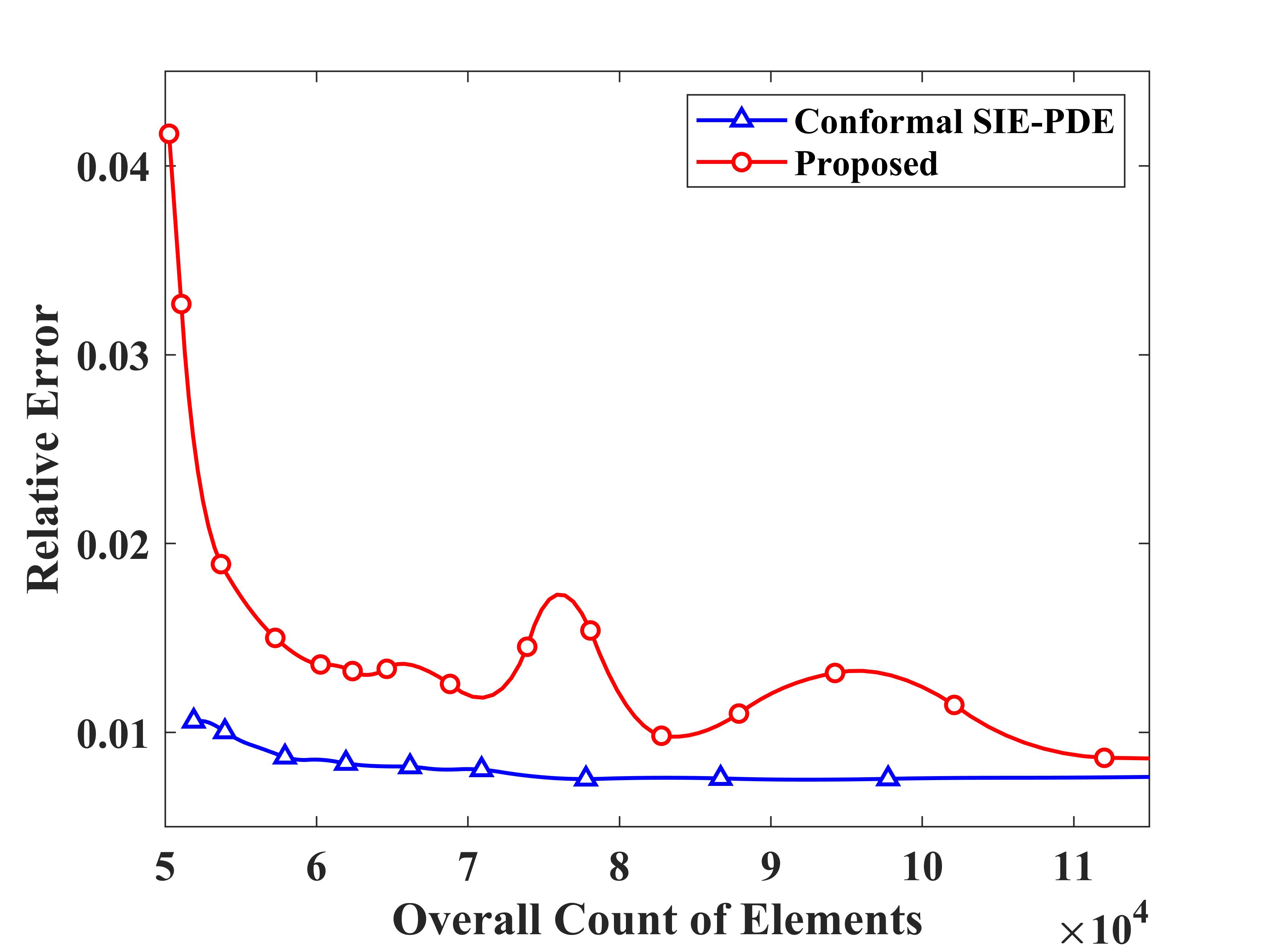}}
		\centering
		\centerline{(b)}
	\end{minipage}
	\caption{Relative error of (a) circular cross section and (b) square cross section, calculated by the conformal hybrid SIE-PDE formulation in [\citen{SunSIE-PDE2022}], and the proposed nonconformal hybrid SIE-PDE formulation versus the overall count of elements.}
	\label{case1_RE}
\end{figure}
	
		Fig. \ref{case1_RCS} (a) and (b) show the RCS for two scenarios, calculated by the COMSOL, the conformal hybrid SIE-PDE formulation in [\citen{SunSIE-PDE2022}], and the proposed nonconformal hybrid SIE-PDE formulation. It can be found that results obtained from the three formulations are all in good agreement. For the proposed formulation, the application of interpolation functions would introduce a certain amount of error, which is closely related to the overall count of elements used in the simulation. Therefore, the relative error (RE) of the RCS obtained from the two hybrid formulations is investigated compared with those from the COMSOL, when the count of elements is increasing. The RE is defined as
		\begin{equation} \label{RE}
		\text{RE} = \frac{{\sum\limits_i {{{\left\| {\text{RCS}^\text{cal} - \text{RCS}^\text{ref}} \right\|}^2}} }}{{\sum\limits_i {{{\left\| \text{RCS}^\text{ref} \right\|}^2}} }},
		\end{equation}
		where $\text{RCS}^\text{cal}$ denotes results calculated from the conformal hybrid formulation in [\citen{SunSIE-PDE2022}], and the proposed nonconformal hybrid formulation, and $\text{RCS}^\text{ref}$ is the reference solutions obtained from the COMSOL.

		As shown in Fig. \ref{case1_RE} (a) and (b), the RE of both formulations gradually decreases as the overall count of elements increases. When the count reaches around 100,000, the relative error can reach less than 1$\%$, which shows comparable convergence properties. More specifically, for the circular object, the RE obtained from both formulations is relatively similar. However, for the square object, the RE of the proposed formulation gradually decreases from the maximum value of 0.0417 to 0.008, while that of the hybrid formulation with conformal meshes gradually varies and gets convergent quite fast. It is expected since the square cross section has non-smooth boundaries. However, the maximum value of the RE from the proposed formulation is only around 0.04, and it decreases fast as the count of elements increases. Therefore, the proposed nonconformal formulation can obtain accurate far fields induced by coated dielectric objects, and achieve similar level of accuracy as the original conformal formulation.
		
\begin{figure*}[t]
			\vspace{-0.3cm}
			\begin{minipage}[h]{0.3\linewidth}
				\centerline{\includegraphics[scale=0.051]{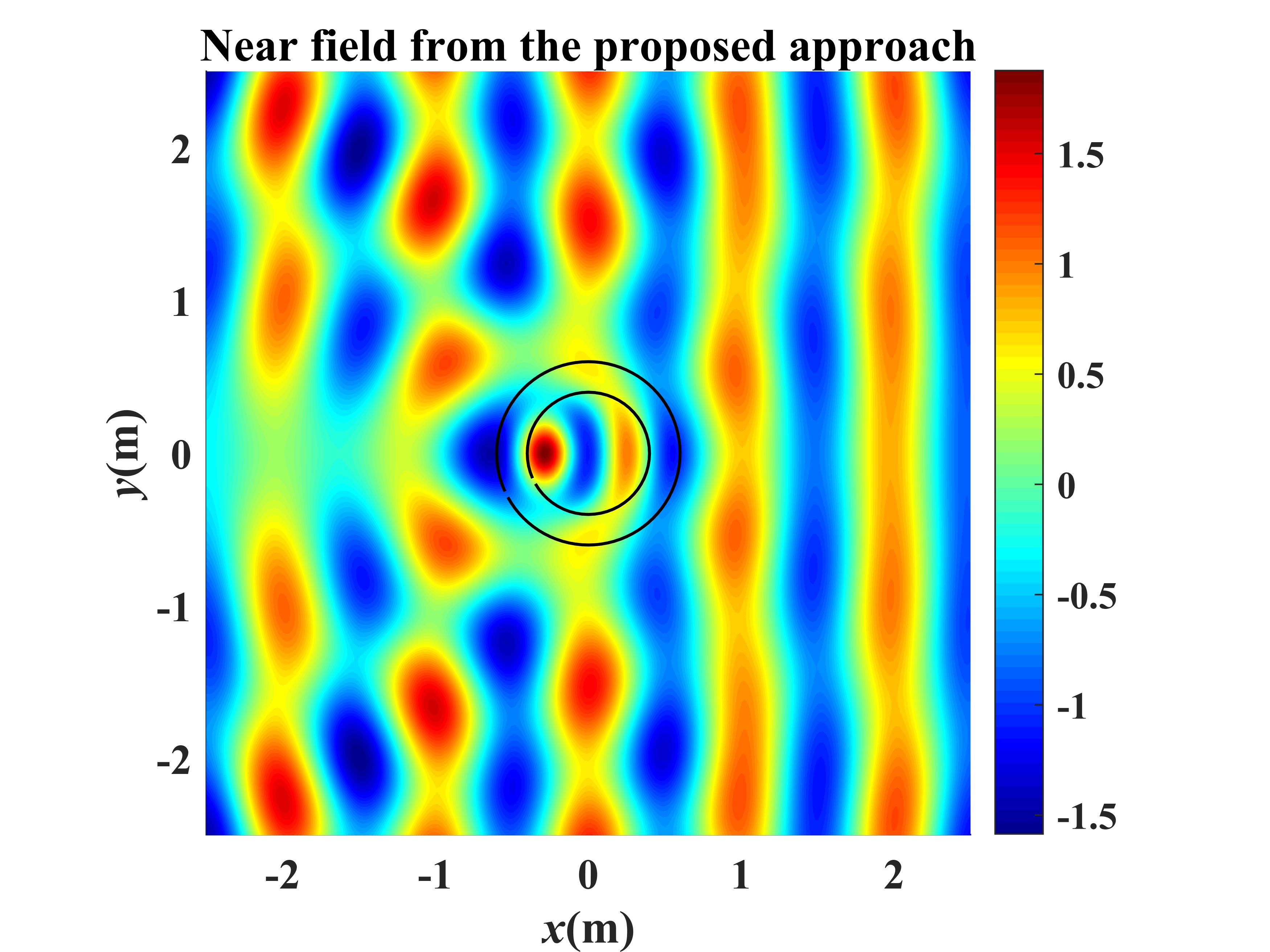}}
				\vspace{-0.05cm}
				\centerline{(a)}
			\end{minipage}
			\hfill
			\begin{minipage}[h]{0.3\linewidth}
				\centerline{\includegraphics[scale=0.051]{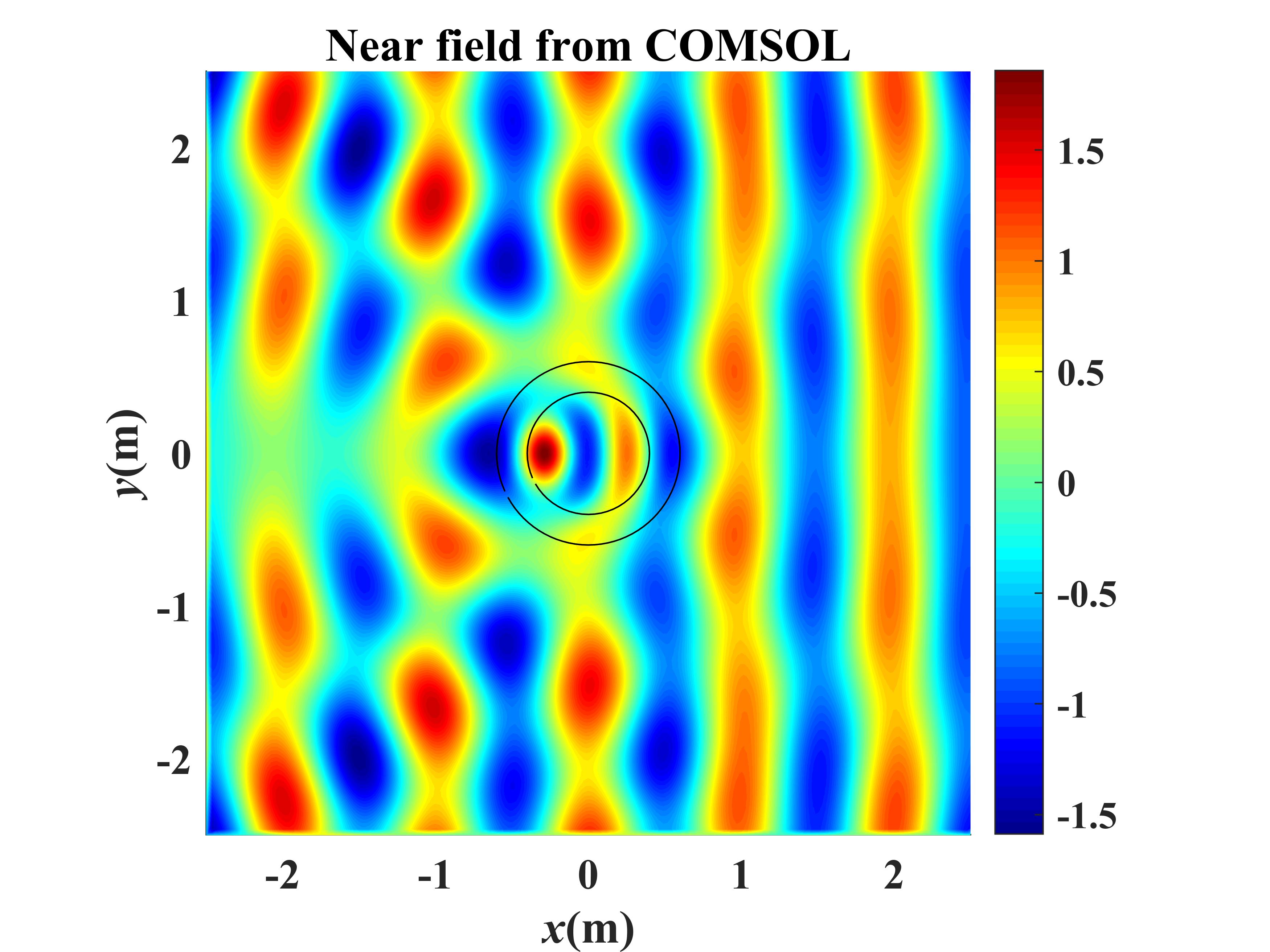}}
				\vspace{-0.05cm}
				\centerline{(b)}
			\end{minipage}
			\hfill
			\begin{minipage}[h]{0.3\linewidth}
				\centerline{\includegraphics[scale=0.051]{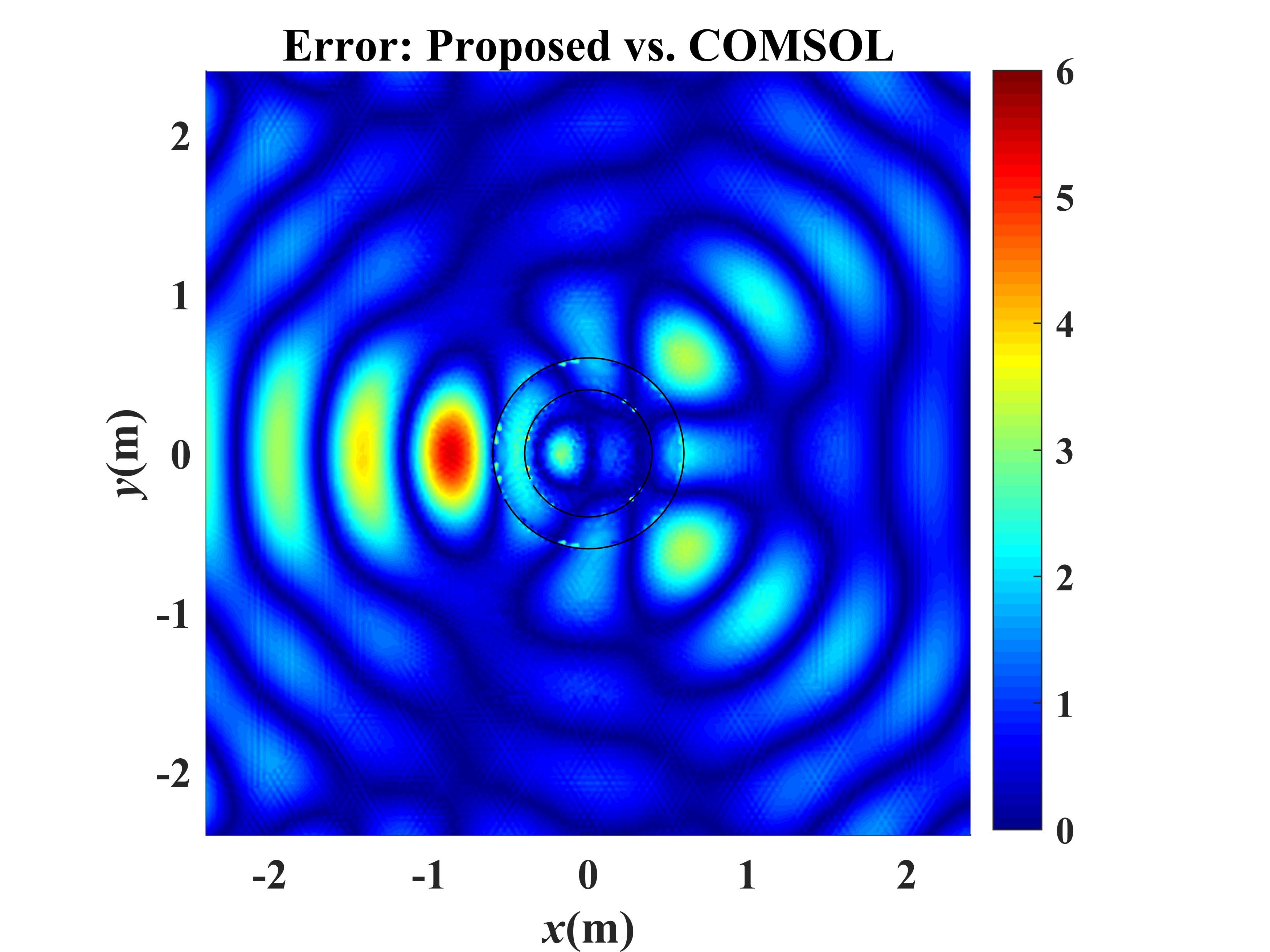}}
				\vspace{-0.05cm}
				\centerline{(c)}
			\end{minipage}
		\vspace{-0.1cm}
			\caption{ Near fields obtained from (a) the proposed nonconformal hybrid formulation, (b) the COMSOL, and (c) the relative error of near fields obtained from the proposed nonconformal hybrid formulation.}
		\label{Nearfield_circle}
		\vspace{-0.3cm}
		\end{figure*}

\begin{figure*}[t]
		\begin{minipage}[h]{0.3\linewidth}
			\centerline{\includegraphics[scale=0.051]{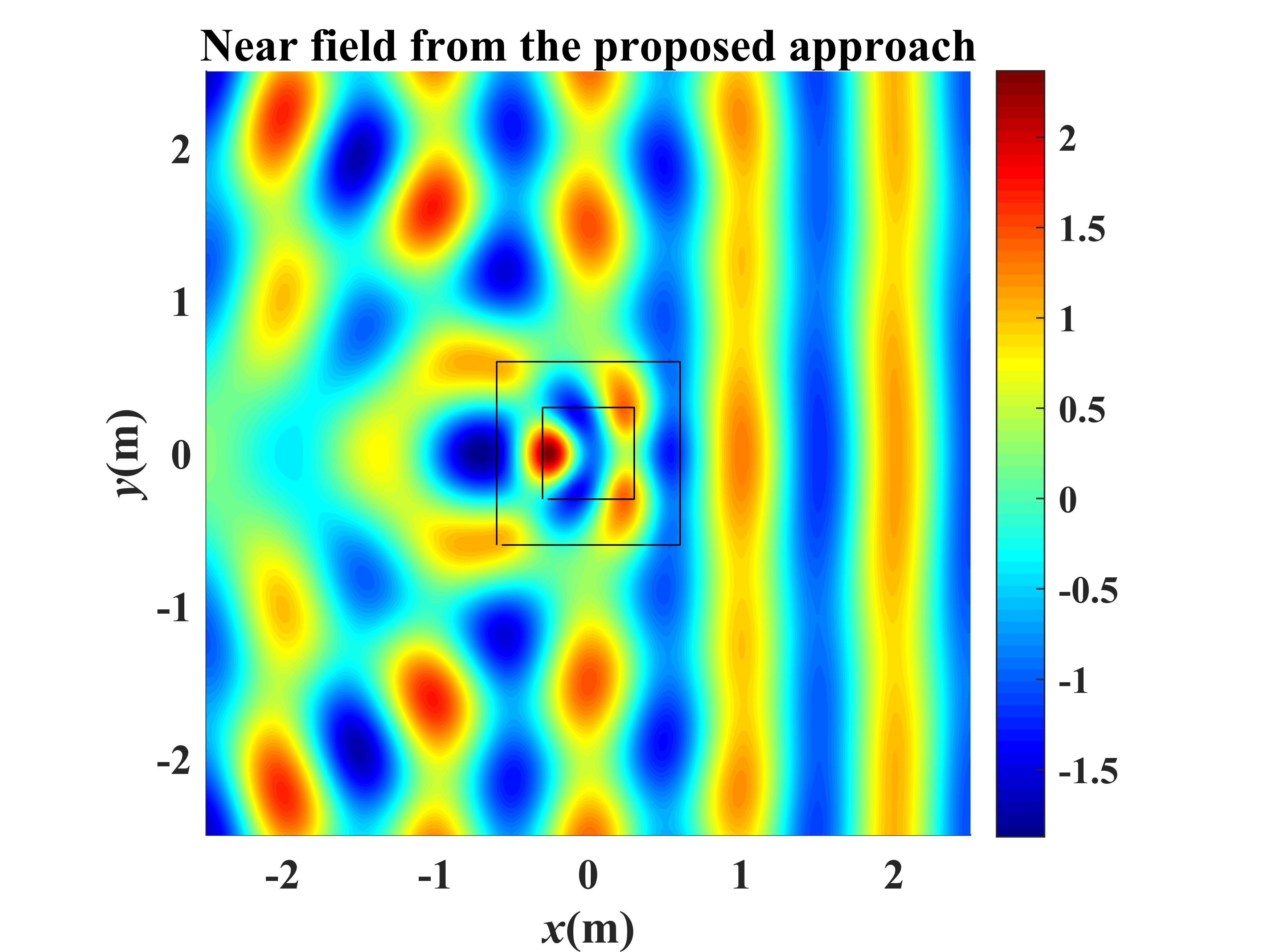}}
			\vspace{-0.05cm}
			\centerline{(a)}
		\end{minipage}
		\hfill
		\begin{minipage}[h]{0.3\linewidth}
			\centerline{\includegraphics[scale=0.051]{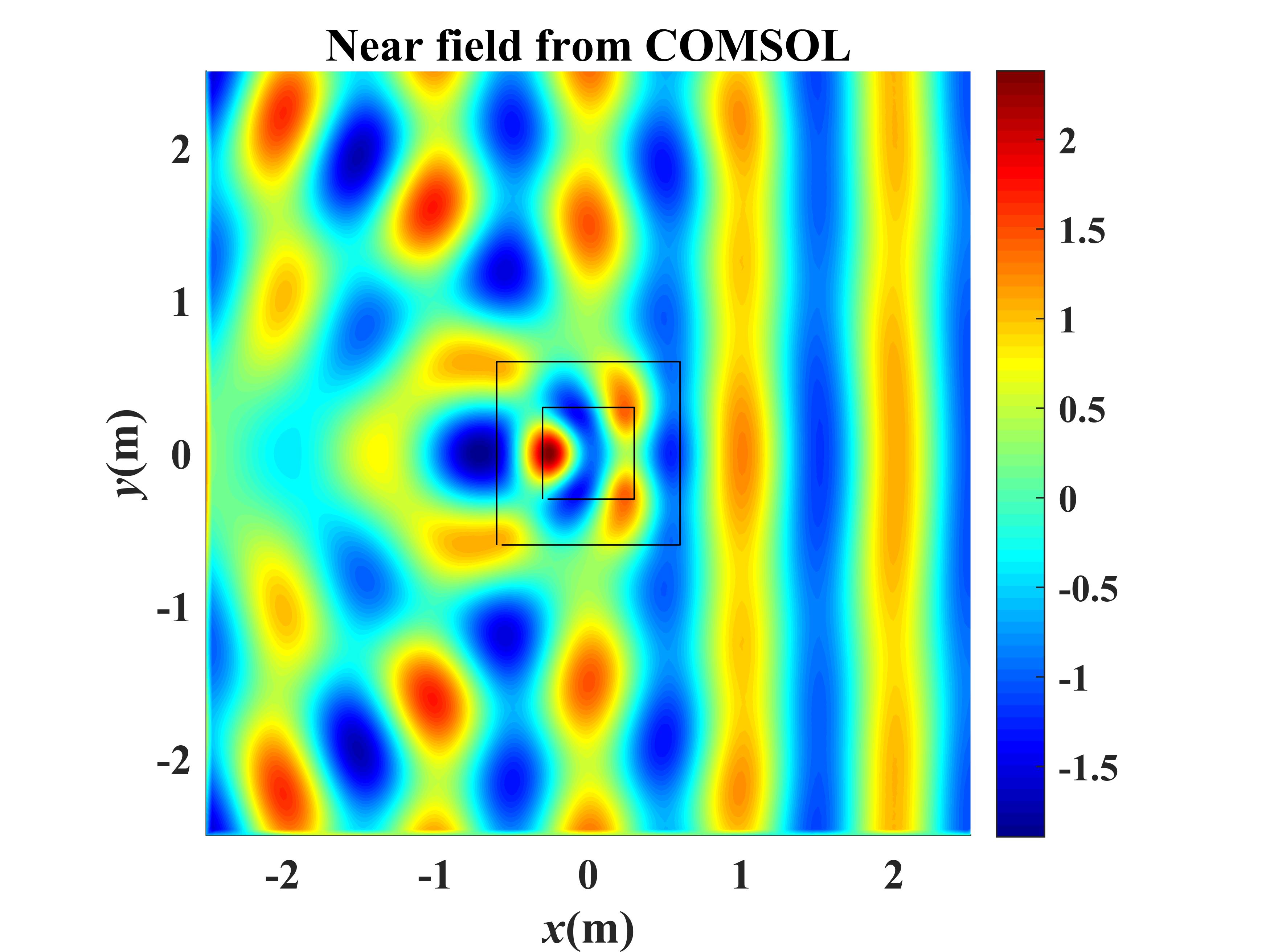}}
			\vspace{-0.05cm}
			\centerline{(b)}
		\end{minipage}
		\hfill
		\begin{minipage}[h]{0.3\linewidth}
			\centerline{\includegraphics[scale=0.051]{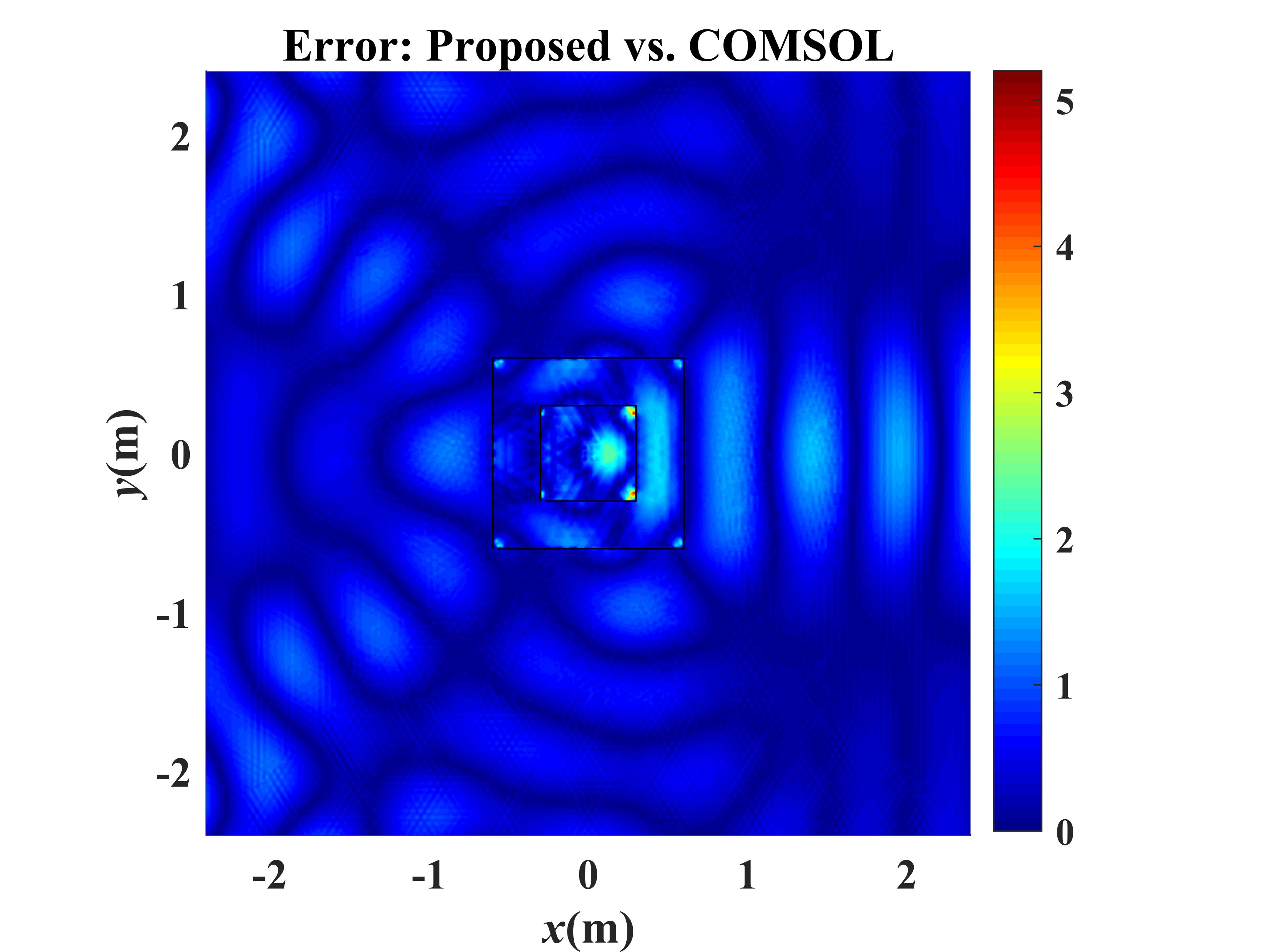}}
			\vspace{-0.05cm}
			\centerline{(c)}
		\end{minipage}
	\vspace{-0.1cm}
		\caption{ Near fields obtained from (a) the proposed nonconformal hybrid formulation, (b) the COMSOL, and (c) the relative error of near fields obtained from the proposed nonconformal hybrid formulation.}
		\label{Nearfield_square}
		\vspace{-0.1cm}
	\end{figure*}

		For further verification, we calculated electric fields near the cylinder through the proposed nonconformal hybrid formulation and the COMSOL in Fig. \ref{Nearfield_circle} (a) and (b), Fig. \ref{Nearfield_square} (a) and (b). It is easy to find that the field patterns obtained from the two formulations are exactly the same and show excellent agreement with each other. To quantitively measure the error, we calculated the relative error of near fields, which is defined as $\left| {{\text{E}^\text{cal}} - {\text{E}^\text{ref}}} \right|/\max \left| {{\text{E}^\text{ref}}} \right|$ , where ${\text{E}^\text{ref}}$ are the solutions obtained from the COMSOL, ${\text{E}^\text{cal}}$ are the solutions obtained from the proposed nonconformal hybrid formulation, and $\max \left| {{\text{E}^\text{ref}}} \right|$ is the maximum magnitude of the reference solutions, which can guarantee the relative error is well defined in the whole computation domain. 
		
	 As shown in Fig. \ref{Nearfield_circle} (c) and Fig. \ref{Nearfield_square} (c), for the circular object, the RE is less than 5$\%$ in most of the computational domain, and for the square object, the RE is no more than 3$\%$ in most regions. A slightly larger error occurs around the boundary of the circular cross section and corners of the square cross section, where the fields charge sharply due to media discontinuity. However, the maximum value is only around 6$\%$ and 5$\%$, respectively. Therefore, near fields of coated dielectric objects can also be accurately calculated through the proposed nonconformal hybrid formulation.	 
	
		\begin{figure}
		\centering
		\vspace{-0.45cm}
		\begin{minipage}[t]{0.5\linewidth}
			\centering
			\centerline{\includegraphics[scale=0.062]{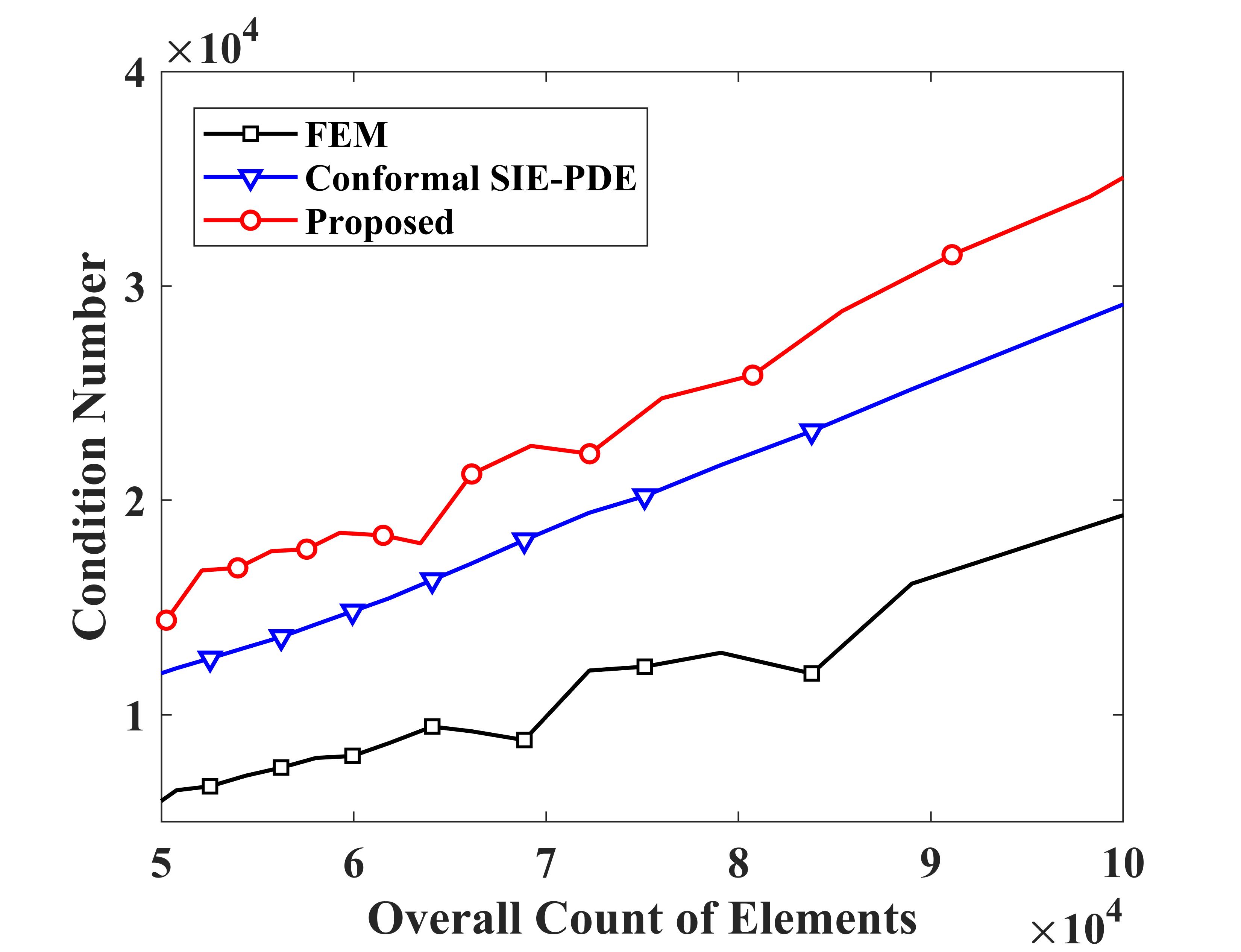}}
			\centering
			\vspace{-0.05cm}
			\centerline{(a)}
		\end{minipage}
		\hfill
		\begin{minipage}[t]{0.5\linewidth}
			\centering
			\centerline{\includegraphics[scale=0.059]{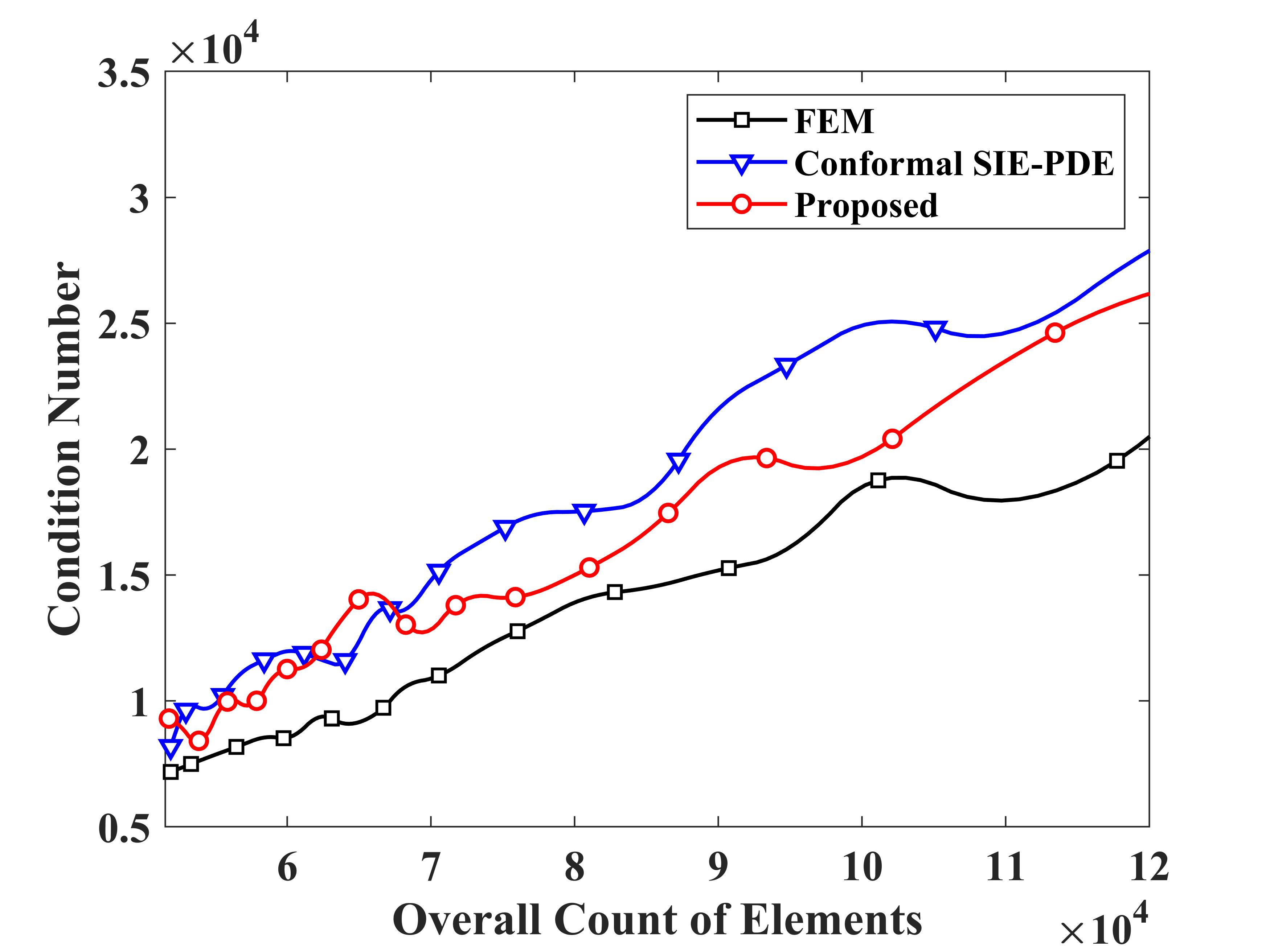}}
			\centering
			\vspace{-0.05cm}
			\centerline{(b)}
		\end{minipage}
		
		\caption{Condition number versus the overall count of elements for dielectric cylinder with (a) circular cross section and (b) square cross section.}
		\label{case1_condition_number}
		\vspace{-0.35cm}
	\end{figure}
	
	In addition, convergence properties of the three formulations in terms of condition numbers are further investigated. The condition number obtained from the three formulations versus the overall count of elements is shown in Fig. \ref{case1_condition_number} (a) and (b). It can be found that condition numbers of the three formulations gradually increase as the count of elements increases. Due to the introduction of the surface equivalent current density, the condition number from the two hybrid formulations is slightly larger than that of the FEM. Besides, the application of interpolation functions also leads to a slightly larger condition number compared to that of the conformal hybrid formulation. However, condition numbers of the three formulations are comparable, which implies that nonconformal meshes have little impact on convergence properties.

	\subsection{Skin Effect Modelling of Cables Placed in the Inhomogeneous Medium}
	If the highly conductive media is involved in the computational domain, it would be challenging to accurately model the resulting skin effect when the frequency is high. The current density crowds around the surface region of highly conductive media, and extremely fine meshes should be used for traditional volumetric formulations, such as the FEM, and the volume integral equation (VIE) formulations. This inevitably leads to large numbers of unknowns and decrease in efficiency.
	
	It is proved that the hybrid SIE-PDE formulation proposed in [\citen{SunSIE-PDE2022}] can mitigate this issue. When the current density with the same level of accuracy is achieved, the SIE-PDE formulation uses fewer unknowns, and less memory and CPU time, which are 29$\%$, 36$\%$, and 7$\%$ of those from the FEM, respectively [\citen{SunSIE-PDE2022}]. The performance improvement is significant. However, since the conformal meshes are used, coarse meshes cannot be applied in the whole region of the PDE domain, especially inside the cable. As a result, the conformal SIE-PDE formulation requires much more unknowns to solve the problem. In addition, it is also inconvenient that every time we make modifications to the SIE domain, meshes for the whole computational domain have to be repeatedly generated. 
	
		 \begin{figure}
		 \vspace{-0.35cm}
		\begin{minipage}[h]{0.22\textwidth}
			\centering
			\centerline{\includegraphics[scale=0.28]{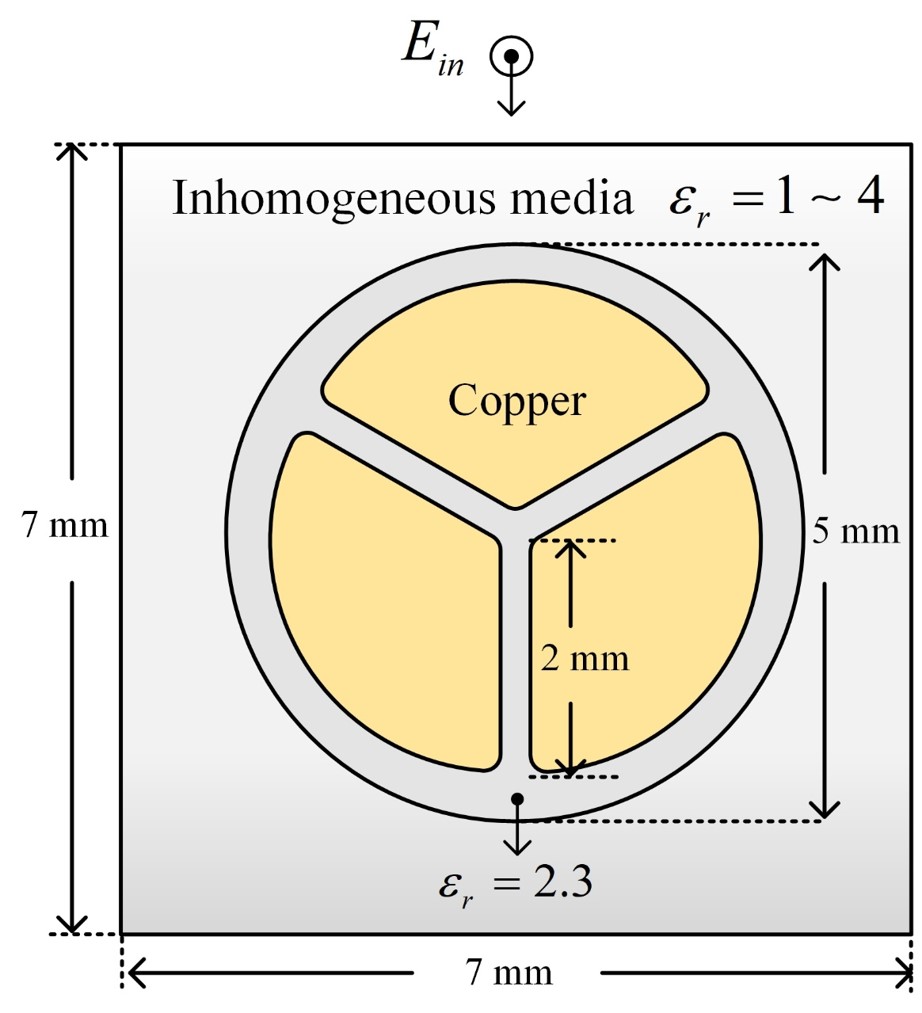}}
			\centering
			\centerline{(a)}
		\end{minipage}
		\begin{minipage}[h]{0.26\textwidth}
			\centering
			\centerline{\includegraphics[scale=0.2815]{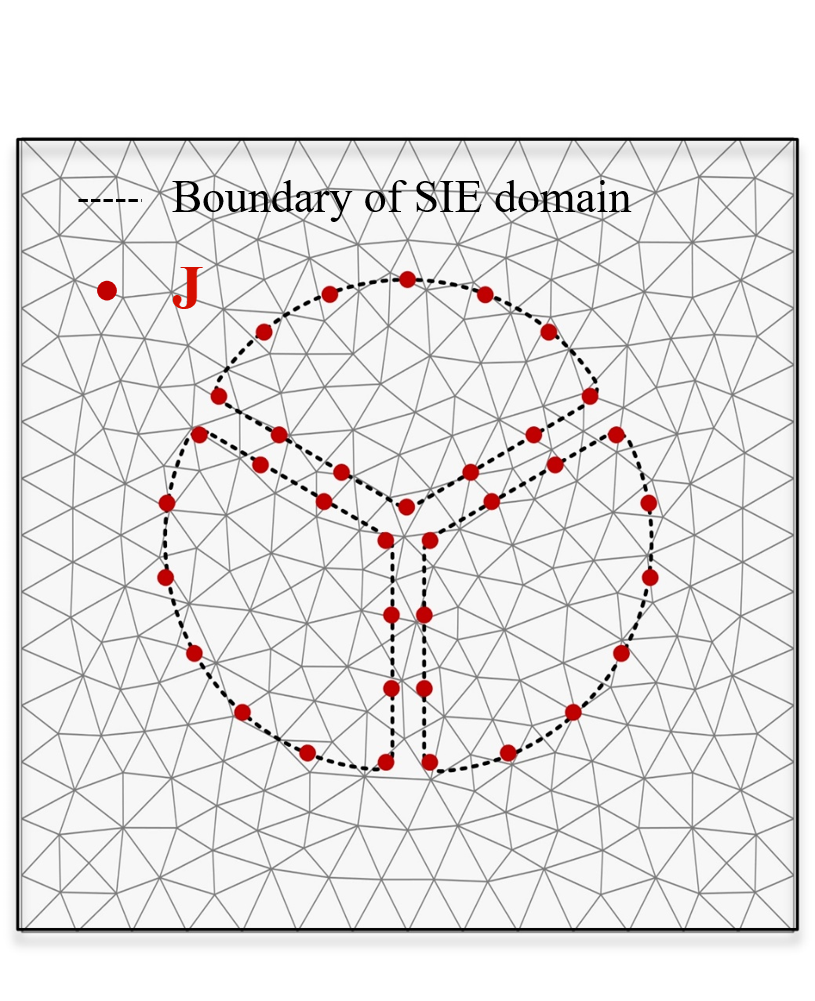}}
			\centering
			\centerline{(b)}
		\end{minipage}
		\caption{(a) Geometrical configurations of a cable placed in the inhomogeneous medium, and (b) nonconformal meshes for the SIE and PDE domains.}
		\label{case3_config}
		\vspace{-0.25cm}
	\end{figure}
	\begin{figure}
		\vspace{-0.25cm}
		\begin{minipage}[t]{0.5\textwidth}
			\centering
			\centerline{\includegraphics[scale=0.063]{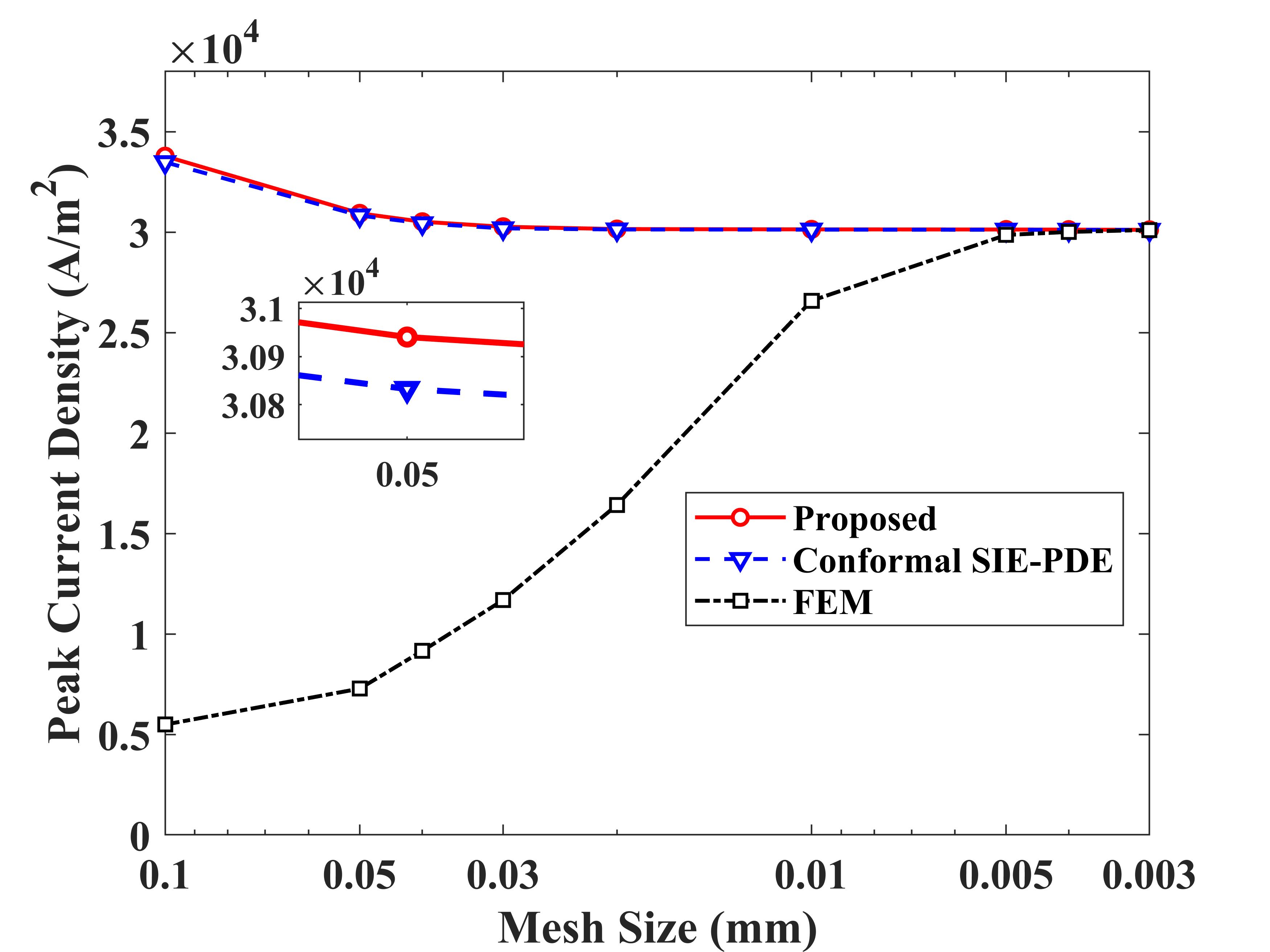}}
			\centering
			\centerline{(a)}
		\end{minipage}
		\begin{minipage}[t]{0.5\textwidth}
			\centering
			\centerline{\includegraphics[scale=0.063]{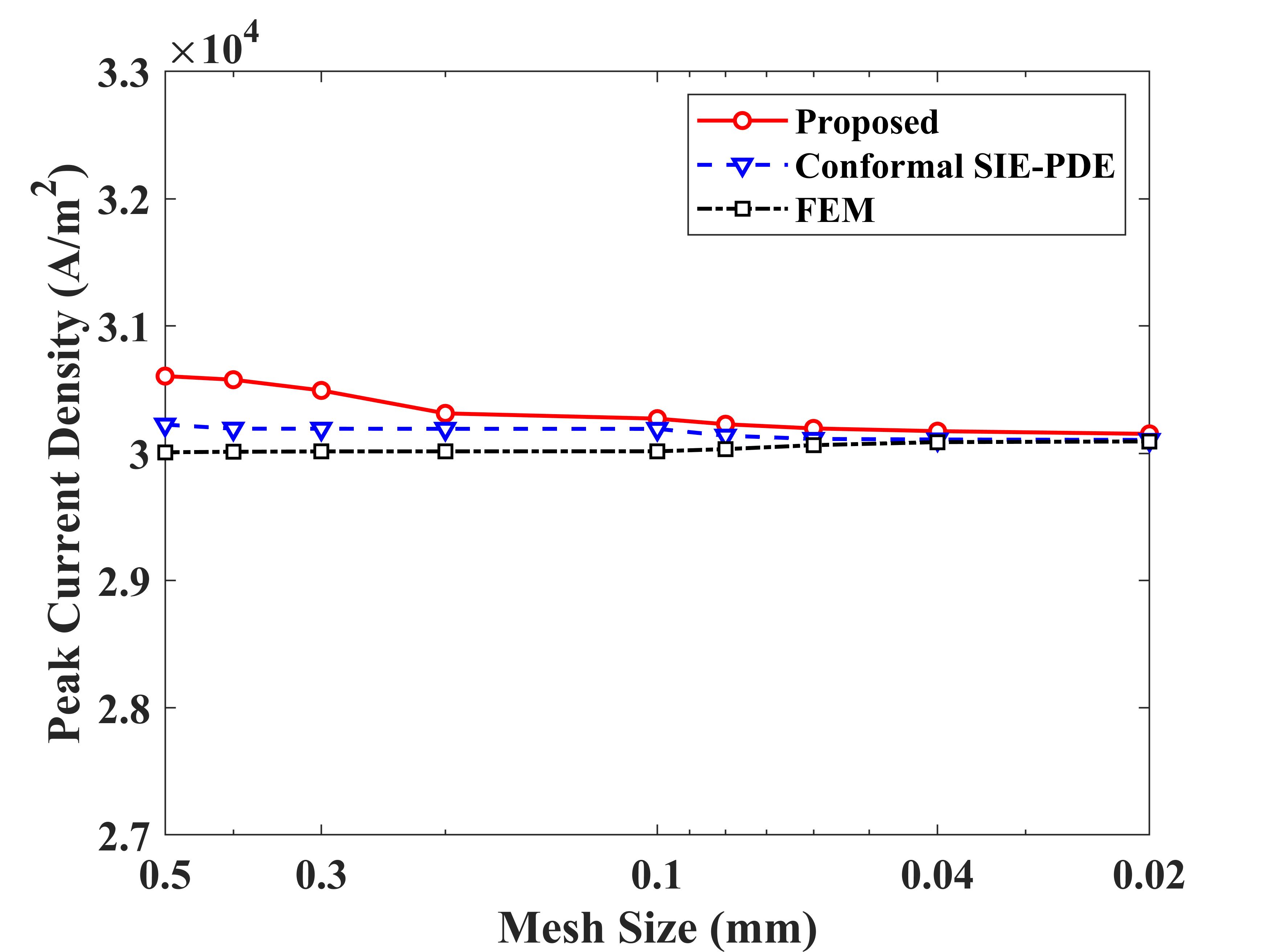}}
			\centering
			\centerline{(b)}
		\end{minipage}
		\caption{The peak current density obtained from the proposed nonconformal hybrid SIE-PDE formulation, the conformal hybrid SIE-PDE formulation in [\citen{SunSIE-PDE2022}] and the FEM in terms of the mesh size of (a) the SIE domain and (b) the PDE domain.}
		\label{case3_current_trend}
		\vspace{-0.4cm}
	\end{figure}
	
	These issues can be properly handled by the proposed hybrid SIE-PDE formulation with nonconformal meshes. The case with a cable placed in the inhomogeneous medium is first considered for accuracy verification. The geometrical configurations are shown in Fig. \ref{case3_config} (a). Three sector-shaped copper conductors are included in the cable with a radius of 2 mm and an angle of 120 degrees, and they are embedded in a dielectric sheath with a radius of 2.5 mm. The relative permittivity of the sheath is 2.3. The cable is embedded in the inhomogeneous medium and its relative permittivity varies linearly along the $y$-axis from 1 to 4. A TM-polarized plane wave with $f = 300$ MHz incidents along the $y$-axis.

	By using the proposed nonconformal SIE-PDE formulation to solve this problem, the whole computational domain is also decomposed into the SIE and PDE domains. The sectors are treated as the SIE domain, which is removed from the computational domain and modeled separately by the SIE formulation. The remaining computational domain is then treated as the PDE domain. The nonconformal meshes for the two domains are shown in Fig. \ref{case3_config} (b). For accuracy verification, results are compared with those from the original conformal SIE-PDE formulation.

	 Since nonconformal meshes are applied, we first verify this impact, and calculate the skin effect in terms of the current density versus mesh sizes in the SIE and PDE domains. Results are compared with those from the conformal SIE-PDE formulation and the traditional FEM with the same size meshes. First, we fix the mesh size for the PDE domain as 0.1 mm, and investigate the impact of mesh sizes of the SIE domain in Fig. \ref{case3_current_trend} (a). The $x$-axis represents mesh sizes in the SIE domain, and the $y$-axis represents the peak amplitude of the current density. It can be found that as mesh sizes decrease, the current density obtained from the FEM gradually increases, and finally is convergent at around 30,000 $A/{m^2}$. However, results from both the conformal and proposed nonconformal SIE-PDE formulations get convergent quite fast, and when the mesh size decreases to 0.05 mm, the current density obtained from the two formulations remains almost unchanged versus mesh sizes. Therefore, the SIE-PDE formulations can accurately model the skin effect with much coarser meshes in the SIE domain, compared with the traditional FEM. By further comparing the two hybrid SIE-PDE formulations, it can be found that the current density obtained from the proposed nonconformal SIE-PDE formulation is slightly larger than that from the conformal SIE-PDE formulation. For example, when the mesh size decreases to 0.05 mm, results from the proposed nonconformal SIE-PDE formulation is 30,940.5 $A/{m^2}$, while that from the conformal SIE-PDE formulation is 30,832.5 $A/{m^2}$. The RE is 0.35 $\%$, which is really small. Then, we investigate the impact of mesh sizes for the PDE domain in Fig. \ref{case3_current_trend} (b). Mesh sizes in the SIE domain are fixed at the values when the current density becomes convergent, which is 0.05 mm for the SIE-PDE formulations, and 0.004 mm for the FEM. It can be found that compared with the other two formulations, for the proposed formulation, mesh sizes in the PDE domain have slightly stronger impact on accuracy. However, it has only quite small influence on the results. When mesh sizes decrease from 0.5 mm to 0.02 mm, the current density from the proposed nonconformal SIE-PDE formulation only varies from 30,606.7$A/{m^2}$ to 30,152.8 $A/{m^2}$. The relative variation is around 1.5 $\%$. Therefore, it can be concluded that mesh sizes in the SIE and PDE domains have small impact on the accuracy of the proposed nonconformal SIE-PDE formulation, and accurate current density can be obtained with coarse meshes both in the SIE and PDE domains.

\begin{figure}
	 	\begin{minipage}[t]{0.5\textwidth}
	 		\centering
	 		\centerline{\includegraphics[scale=0.63]{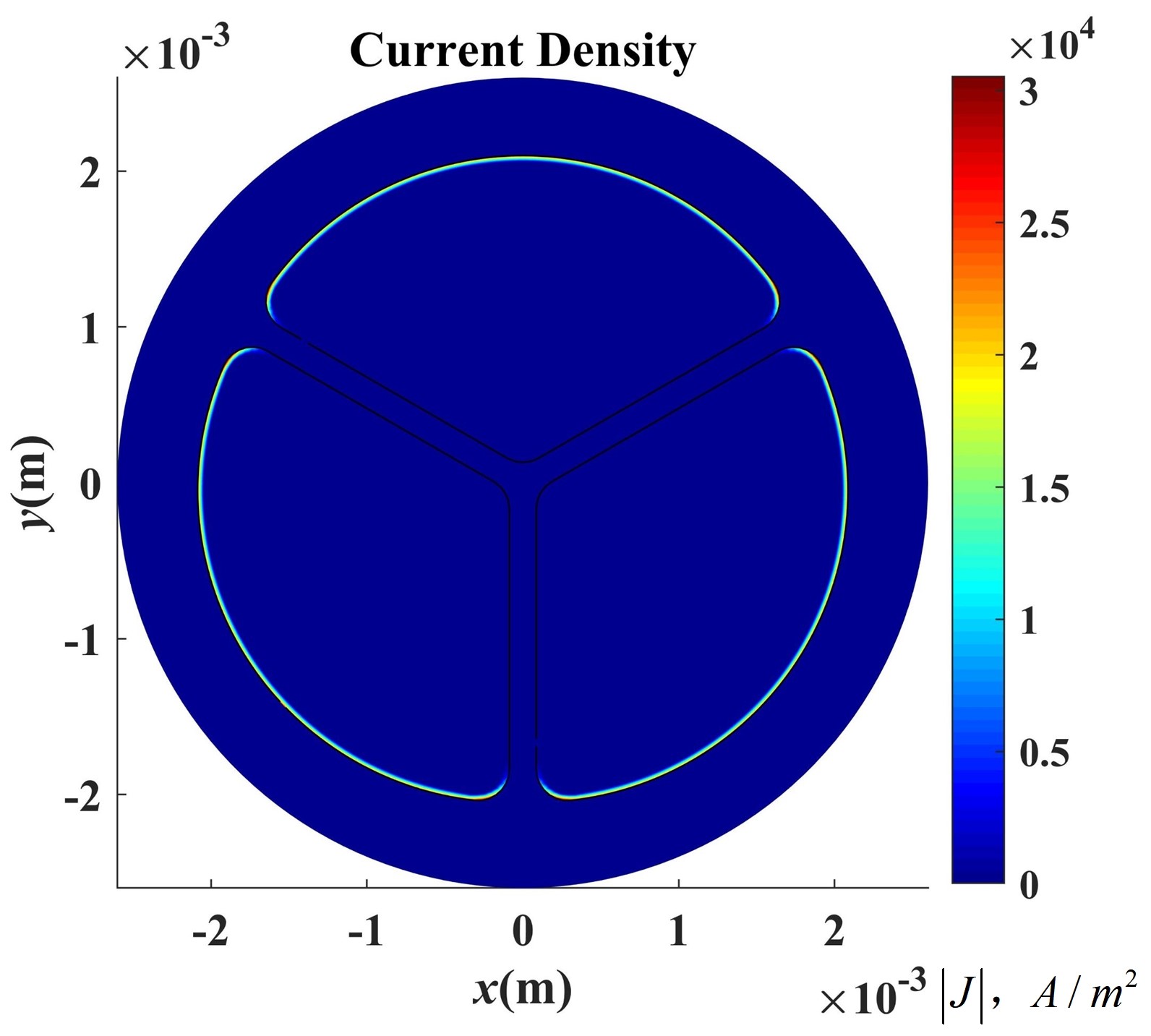}}
	 		\centering
	 		\centerline{(a)}
	 	\end{minipage}
	 	\begin{minipage}[t]{0.5\textwidth}
	 		\centering
	 		\centerline{\includegraphics[scale=0.63]{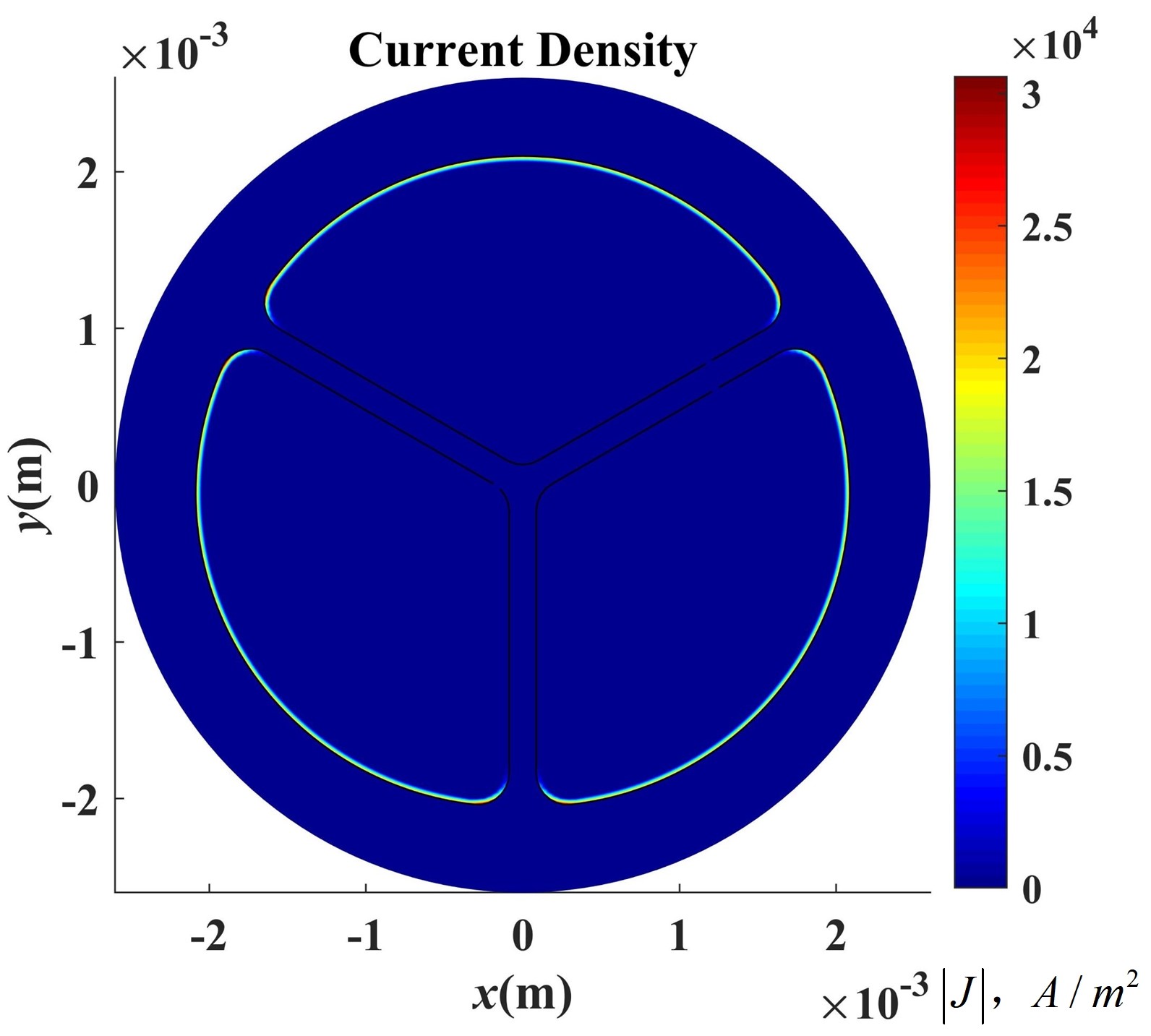}}
	 		\centering
	 		\centerline{(b)}
	 	\end{minipage}
	 	\caption{The current density obtained from (a) the proposed nonconformal hybrid SIE-PDE formulation, and (b) the conformal hybrid SIE-PDE formulation in [\citen{SunSIE-PDE2022}].}
	 	\label{case3_current_density}
	 \end{figure}

	 Fig. \ref{case3_current_density} (a) and (b) show the convergent results of the current density obtained from both nonconformal and conformal SIE-PDE formulation. The mesh configurations are set as the values when the current density gets convergent, which is 0.05 mm for the SIE domain, and 0.2 mm for the PDE domain. It is easy to find that the patterns of the current density are almost the same. The peak value of the current density is 30,313.9 $A/{m^2}$ for the proposed nonconformal SIE-PDE formulation, compared with 30,192.6 $A/{m^2}$ for the conformal SIE-PDE formulation. The relative error is around 0.4$\%$, which implies that both formulations can accurately capture the skin effect in terms of the current density.
	 
\begin{figure}[t]
		\centering
		\includegraphics[width=0.5\textwidth]{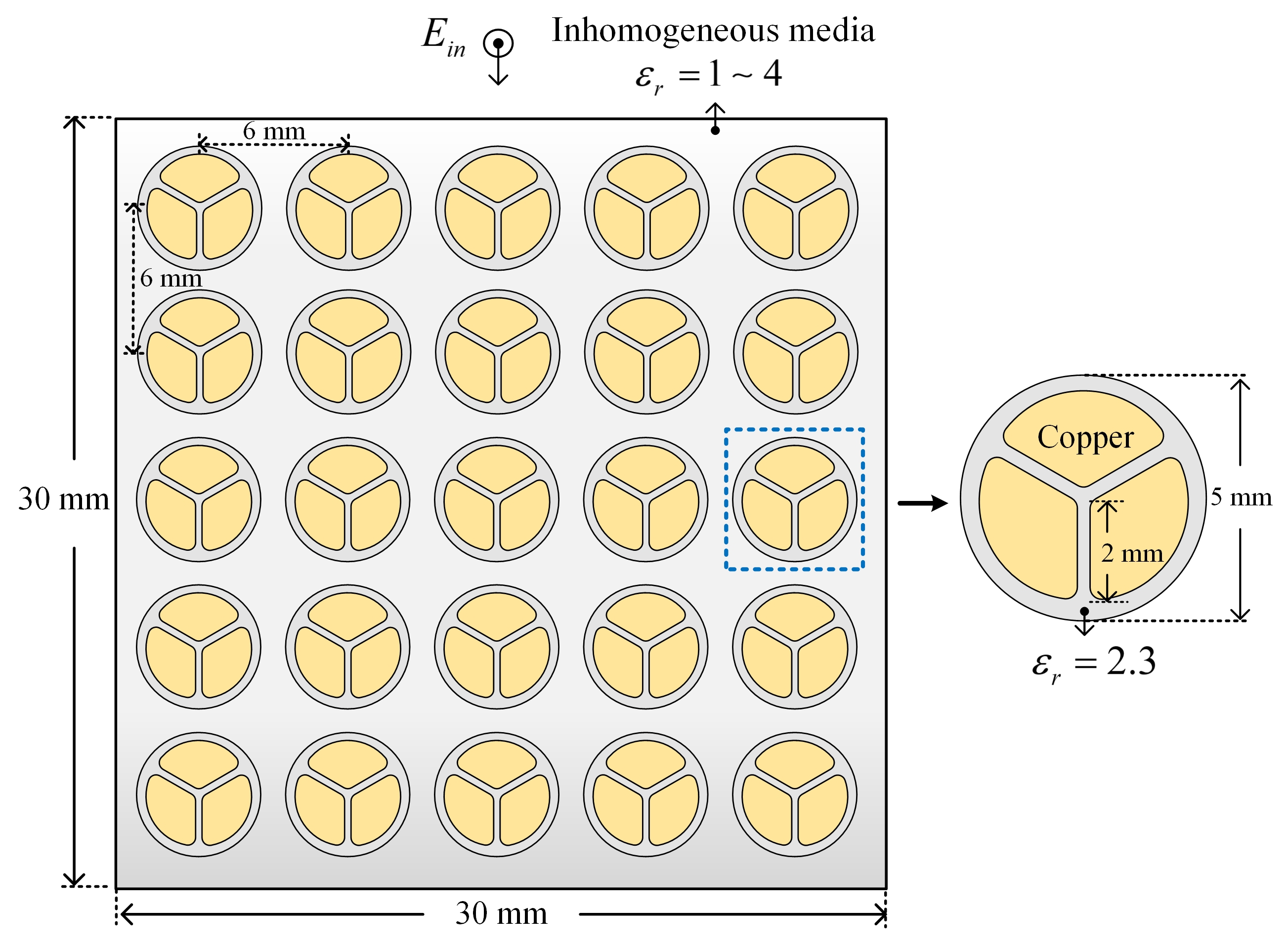}
		\caption{Geometrical configurations of an array with 25 cables placed in the inhomogeneous medium.}
		\vspace{-0.2cm}
		\label{case4_Geo} 
	\end{figure}

	 Based on the numerical verification above, we move on to a large problem, and further verify the efficiency of the proposed formulation. A muti-cable system consisting of an array with 25 cables is considered, and its geometric configurations are shown in Fig. \ref{case4_Geo}. The material properties are the same as the single cable case. The array is fully embedded in the inhomogeneous medium, and its permittivity also varies linearly along the $y$-axis from 1 to 4. Through the proposed nonconformal SIE-PDE formulation, the whole computational domain is decomposed into the SIE and PDE domains as well, and for each cable, the nonconformal meshes are similar to the single cable case as shown in Fig. \ref{case3_config} (b). It should be noted that since all cables share the same geometrical and material configurations, the related DSAO only needs to be generated once. Results are compared with those from the conformal SIE-PDE formulation and the FEM. The mesh configurations are set as the values when the current density gets convergent. For the SIE-PDE formulations, they are 0.05 mm for the SIE domain, and 0.2 mm for the PDE domain, while for the FEM, mesh sizes are 0.004 mm and 0.2 mm, respectively. 

\begin{figure*}[t]
	\includegraphics[width=1.0\textwidth]{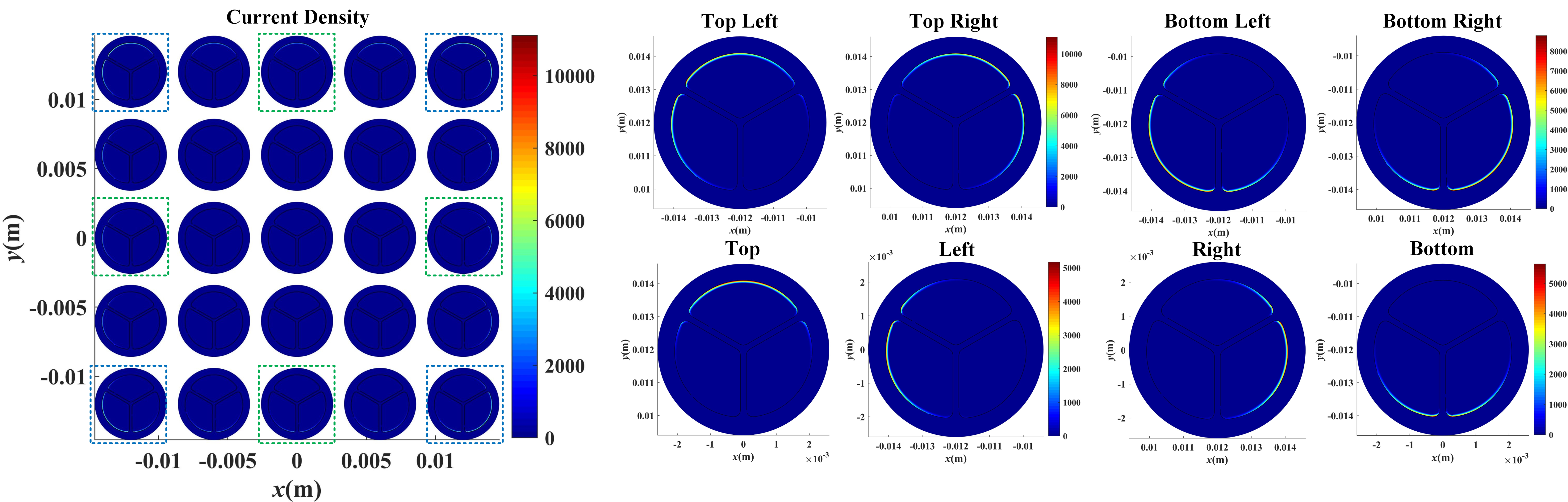}
	\caption{The current density obtained from the proposed nonconformal hybrid SIE-PDE formulation.}
	\label{case4_result} 
\end{figure*}

	Fig. \ref{case4_result} shows the current density obtained from the proposed nonconformal SIE-PDE formulation. Since there is no current density in the inhomogeneous medium, it can be ignored and only the results inside cables are presented. The zoom-in views of the cables with typical patterns are also shown in Fig. \ref{case4_result}. Among them, four are at the corners of the array, and the other four are at the middle of the edges. They are marked with blue and green boxes in the results of the whole array, respectively. Table I shows the comparison of the computational consumption of the SIE-PDE formulations and the FEM when the current density with the same level of accuracy is obtained. It should be noted that since our in-house codes have been optimized, the time consuming of the FEM and the conformal SIE-PDE formulation is much reduced compared with that listed in [\citen{SunSIE-PDE2022}]. As shown in the second row of Table I, the peak value of the current density is 11,154.9 $A/{m^2}$ for the proposed nonconformal SIE-PDE formulation, 11,107.0 $A/{m^2}$ for the conformal SIE-PDE
	formulation, and 10,920.9 $A/{m^2}$ for the FEM. Therefore, all the three formulations achieve the same level of accuracy. Then, we calculated the overall unknowns used for the three formulations. Since nonconformal meshes are used in the proposed formulation, its unknowns only reside in the PDE domain where much coarser meshes are used, compared with those for the other two formulations. Therefore, the proposed nonconformal SIE-PDE formulation needs the fewest unknowns to solve this problem. The overall count of unknowns is 69,360, which is 23.1$\%$ and 0.2$\%$, respectively, of those in the conformal SIE-PDE formulation and the FEM. Furthermore, as shown in the fourth row of Table I, the reduction in the memory consumption is also significant for the proposed nonconformal SIE-PDE formulation, which is only 22.9$\%$ and 0.2$\%$ of that of the conformal SIE-PDE formulation and the FEM, respectively. The time cost for the three formulations is shown from the fifth to seventh rows in Table I. The time cost includes generation for $\mathbb{Y}_s$, and matrices filling and solving. Among the three formulations, the time cost is the most for the FEM to solve this problem, which is 776.0 seconds, since its large count of unknowns leads to a relatively large-scale matrix, and needs much more time for the matrices filling and solving. Compared with the FEM, both hybrid SIE-PDE formulations use much less CPU time in this simulation, which is 6.7 seconds for the conformal SIE-PDE formulation, and 4.1 seconds for the nonconformal SIE-PDE formulation. They are only 0.8$\%$ and 0.5$\%$ of that from the FEM, respectively. Further comparing the two hybrid SIE-PDE formulations, it can be find from the last two rows of Table I that the two formulations share almost the same time cost for the generation of $\mathbb{Y}_s$, while the proposed nonconformal formulation needs much less CPU time for the matrices filling and solving, which is only 30.3$\%$ of that for the conformal formulation. It is expected that with nonconformal meshes, much coarser meshes are used in the computational domain, along with much fewer unknowns compared with conformal meshes. It greatly reduces the computational consumptions to complete this simulation. 

\vspace{0.2cm}

	\begin{table}
	\centering
	\caption{Comparison of computational costs for the FEM, the conformal SIE-PDE formulation, and the proposed nonconformal SIE-PDE formulation}\label{table1}
	\vspace{0.2cm}
		\resizebox{8.8cm}{!}{
	\begin{tabular}{l|c|c|c}
		\hline
		\hline
		\textbf{ } &\multirow{2}{*}{\textbf{FEM}}\   & \multirow{2}{*}{\textbf{Conformal}} & \multirow{2}{*}{{\textbf{Proposed}}}\\
		{} & & &\\ 
		\hline
		\hline
		Peak Value of Current         &\multirow{2}{*}{10,920.9}  & \multirow{2}{*}{11,107.0}         &\multirow{2}{*}{11,154.9}      \\
		Density [$A/m^2$] & & &\\
		\hline
		Overall Counts of      &\multirow{2}{*}{29,077,493}   & \multirow{2}{*}{300,889} & \multirow{2}{*}{69,360}    \\
		Unknowns & & &\\ 
		\hline
		Memory           &\multirow{2}{*}{50,626.0}     & \multirow{2}{*}{504.0}   &\multirow{2}{*}{115.7}     \\
		Consumption [MB] & & & \\
		\hline
		\multirow{2}{*}{\textbf{Total Time } [s]} &\multirow{2}{*}{776.0} &\multirow{2}{*}{6.7}   & \multirow{2}{*}{4.1}  \\
		{} & & &\\
		\hline
		Time for $\mathbb{Y}_s$ &\multirow{2}{*}{--}    &\multirow{2}{*}{3.4}  & \multirow{2}{*}{3.1}      \\
		Generation [s] & & &\\
		\hline
		Time for Matrices       &\multirow{2}{*}{776.0}  &\multirow{2}{*}{3.3}  &\multirow{2}{*}{1.0}       \\  
		Filling and Solving [s] & & &\\
		\hline 
		\hline 
	\end{tabular}
		}
\end{table}
\vspace{0.1cm}

	To sum up, the proposed hybrid SIE-PDE formulation with nonconformal meshes can accurately and efficiently capture the skin effect in terms of the current density. The application of the nonconformal meshes allows for much coarser meshes in the PDE domain, which further reduces the count of unknowns, CPU time and memory consumption. Besides, if modifications are made to the cables, only the meshes for the SIE domains need to be regenerated, which greatly improves the flexibility.


	\subsection{Discussion}
	From the above numerical cases, the computational accuracy and efficiency of the proposed hybrid SIE-PDE formulation with nonconformal meshes has been detailed presented. It should be noted that when the SIE domain occupies a large proportion of the whole computational domain, the count of unknowns can be greatly reduced with the proposed formulation, since the overall unknowns only reside in the PDE domain where much coarser meshes can be used. Therefore, significant improvements can be achieved in computational efficiency with much fewer unknowns, and less CPU time and memory consumption, in comparison with the SIE-PDE formulation proposed in [\citen{SunSIE-PDE2022}]. This advantage makes the proposed formulation more attractive in solving multiscale and complex structures with computationally challenging media.
	
	When the SIE domain scales larger, the proposed formulation can still greatly reduce the count of unknowns. However, the inversion of the resulting dense matrices would increase the computational burden which may degrade the efficiency. This issue can be mitigated with the nonconformal SIE-DDMs [\citen{ZhuSS-SIE2022}]. By decomposing the large SIE domain into subdomains, the dimensions of the matrices requiring to be inverted can be much reduced. In addition, since in our in-house codes, all the inversion of matrices is done by the direct solver pre-defined in the Matlab, it would be computationally expensive with the problems scaling. This can also be mitigated by using iterative algorithms along with acceleration methods, such as MLFMA [\citen{SongMLFMA1997}], pFFT [\citen{PhilipspFFT1997}], and AIM [\citen{BleszynskiAIM1996}].
	
	The proposed hybrid formulation with nonconformal meshes can be extended into 3D scenarios by exchanging the 2D nodal-based linear basis function with 3D edge-based vector basis function [\citen{Jin2015FEM}, Ch. 8, pp. 420-435]. Then to support nonconformal meshes, these vector basis functions can also be served as interpolation functions and used to construct the connection matrix $\mathbb{T}$.

	\section{Conclusion}
	In this paper, an efficient and simple SIE-PDE formulation with nonconformal meshes for TM electromagnetic analysis is developed. In this formulation, an equivalent model with only the electric current density is first constructed in the SIE domain, and then is coupled into the inhomogeneous Helmholtz equation in the PDE domain. By carefully constructed appropriate basis and interpolation functions, the two domains are coupled without additional boundary condition requirements, and nonconformal meshes are supported through a connection matrix $\mathbb{T}$. The proposed nonconformal SIE-PDE formulation shows many significant advantages over other existing hybrid formulations, like no extra boundary conditions, easy implementation, and nonconformal mesh support. These merits are quite appealing in solving challenging electromagnetic problems, like scattering from multiscale structures, modeling and parameter extraction of complex integrated circuits with computationally challenging media. Numerical results also demonstrate that the proposed SIE-PDE formulation with nonconformal meshes can significantly improve the performances in terms of the numbers of unknowns, CPU time and memory consumption over the conformal SIE-PDE formulation and the FEM. 
	
	Extension of current work into 3D scenarios is in progress, and we will report more results on this topic in the future.

\end{document}